\documentclass[aps,pre,epsf,superscriptaddress,amsmath,amssymb,amsfonts,twocolumn,showpacs]{revtex4-1}

\usepackage{graphicx}
\usepackage{epsfig}
\usepackage{dcolumn}
\usepackage{bm}
\usepackage{braket}
\usepackage{amsmath}
\usepackage{color}

\begin{document}

\title{Mode coupling of interaction quenched ultracold few-boson\\ ensembles in periodically driven lattices}

\author{S.I. Mistakidis}
\affiliation{Zentrum f\"{u}r Optische Quantentechnologien,
Universit\"{a}t Hamburg, Luruper Chaussee 149, 22761 Hamburg,
Germany}
\author{P. Schmelcher}
\affiliation{Zentrum f\"{u}r Optische Quantentechnologien,
Universit\"{a}t Hamburg, Luruper Chaussee 149, 22761 Hamburg,
Germany} \affiliation{The Hamburg Centre for Ultrafast Imaging,
Universit\"{a}t Hamburg, Luruper Chaussee 149, 22761 Hamburg,
Germany}

\date{\today}

\begin{abstract}
The out-of-equilibrium dynamics of interaction quenched finite ultracold bosonic ensembles in periodically driven
one-dimensional optical lattices is investigated.
It is shown that periodic driving enforces the bosons in the outer wells of the finite lattice to exhibit  
out-of-phase dipole-like modes, while in the central well 
the atomic cloud experiences a local breathing mode. The dynamical 
behavior is investigated with varying driving frequency, revealing a resonant-like behavior 
of the intra-well dynamics. An interaction 
quench in the periodically driven lattice gives rise to admixtures of different excitations in the outer wells, 
an enhanced breathing in the center and 
an amplification of the tunneling dynamics. We observe then multiple resonances 
between the inter- and intra-well dynamics at different quench amplitudes, with  
the position of the resonances being tunable via the driving frequency. 
Our results pave the way for future investigations on the use of combined driving protocols in order 
to excite different inter- and intra-well modes and to subsequently control them.\\

\end{abstract}

\maketitle

\section{Introduction}

Ultracold atoms in optical lattices offer an ideal platform for simulating certain problems of condensed matter physics 
and constitute many-body systems exhibiting a diversity of physical 
phenomena. In particular, the understanding of the non-equilibrium dynamics of strongly correlated many-body systems 
in optical lattices is currently 
one of the most challenging problems for both theory and experiment. This dynamics is typically triggered by an   
external periodic driving \cite{Goldman,Goldman1,Morsch1,Bloch} or an instantaneous change (quench) of a Hamiltonian 
parameter \cite{Polkovnikov}. Remarkable dynamical phenomena employing 
a periodic driving \cite{Goldman,Goldman1} of the optical lattice include Bloch-oscillations \cite{Dahan,Morsch,Hartmann}, 
the realization of the superfluid to Mott insulator phase transition \cite{Eckardt}, topological states of matter \cite{Zheng},  
artificial gauge fields \cite{Struck}, the realization of ferromagnetic domains \cite{Parker,Choudhury} and even applications to quantum 
computation \cite{Schneider1}. On the other 
hand, quench dynamics enables us to explore among others the light-cone-effect in the spreading of 
correlations \cite{Cheneau,Natu}, the Kibble-Zurek mechanism \cite{Zurek,Chen1} or the 
question of thermalization \cite{Rigol,Altman}. Driving or quenches can also be used    
in order to generate energetically low-lying collective modes, such as the dipole \cite{Kohn, Bonitz} or the  
breathing mode \cite{Abraham,Bauch1,Abraham1,Schmitz,Peotta}. In general, 
a sudden displacement or a periodic shaking of the external trap induces a dipole oscillation of the atomic cloud, while a 
quench on the frequency of the trap excites a breathing mode of the cloud. These modes constitute a main probe both 
for theoretical investigations, to understand and interpret the non-equilibrium dynamics, and for experiments, as they can be used in order to measure 
key quantities of trapped many-body systems \cite{Abraham}.

Recently, increasing effort has been devoted to control the atomic motion in optical lattices by 
subjecting them to a time-periodic external driving \cite{Lignier,Sias,Haller,Chen} and investigating the optimal 
driving protocol \cite{Rosi,Brif,Brif1}. In this direction, it is important to carefully explore and design the relevant driving 
protocol to transfer the energy to the desired final degrees of freedom.  
To trigger or even control a certain type of (collective) modes of the dynamics, widely used techniques in the
literature constitute either the periodic driving of the lattice potential, e.g. a lattice shaking, or 
a quench of a parameter of the system, e.g. a lattice amplitude quench or an interaction quench.
In the former case a tunable local dipole mode and a resonant intra-well dynamics were recently explored by shaking an optical lattice \cite{Mistakidis2}.
On the other hand, in the latter case it has been shown \cite{Mistakidis} that a
sudden increase of the inter-particle repulsion in a non-driven lattice induces a rich inter-well as well as intra-well dynamics which can be coupled and consequently
mixed for certain quench amplitudes. However, for decreasing repulsive forces \cite{Mistakidis1} the accessible
inter-well tunneling channels are much fewer compared to the excited intra-well modes, and
in particular no resonant dynamics can be observed.
From the above analysis it becomes evident that a crucial ingredient for the design and further
control of the dynamics is the choice of the driving protocol of the system: By using different
driving schemes, different types of excited modes are induced, i.e. different energetical
channels can be triggered.
In this direction, an intriguing question is how a combination of periodic driving and interaction quenches can be used to steer the dynamics 
of the system and as a consequence the coupling of the inter-well and intra-well modes. Such an investigation will, among others, permit us 
to gain a deeper understanding of the underlying microscopic mechanisms,
and will allow us to activate certain energy channels by using specific driving protocols for the control of the
different processes.

In the spirit of the above-posed question we investigate in the
present work the quantum dynamics of interaction quenched few-boson ensembles trapped in
periodically driven finite optical lattices.
Concerning the periodic driving, a vibration of the optical lattice is employed.
This scheme, in contrast to shaking, induces out-of-phase dipole modes among the outer wells
and a local breathing mode in the central well of the finite lattice. 
We cover the dynamics of the periodically driven lattice with varying driving frequency
in the complete range from adiabatic to high frequency driving.
In particular we observe for the intermediate driving frequency regime, being intractable by current state
of the art analytical methods \cite{Goldman,Goldman1}, a resonant-like behavior of the intra-well dynamics.
This resonance is accompanied by a rich excitation spectrum and an enhanced inter-well tunneling as compared 
to adiabatic or high intensity driving and it is mainly of single particle character. Indeed, it survives 
upon increasing interaction obtaining faint additional features the most remarkable being the co-tunneling of an atom pair \cite{Chen,Folling}.
To induce a correlated many-body 
dynamics we employ an interaction quench on top of the driven lattice,
thus opening energetically higher inter-well and intra-well channels. As a consequence the inter-well tunneling is amplified even for adiabatic driving 
and admixtures of excitations possessing breathing-like and dipole-like components are generated. Remarkably enough, as a function of the quench amplitude, 
the system experiences multiple resonances between the inter- and intra-well dynamics. This observation indicates the high degree of controlability of the 
system especially for the excited modes
under such a combination of driving protocols and it is arguably one of our central results. To the best of our knowledge, this multifold mode coupling behavior 
unraveled with a composite
driving protocol has never been reported before. Moreover, the position of the above mentioned resonances is tunable via the driving frequency allowing for 
further control of the mode coupling in optical lattices. Finally, the realization of intensified loss of coherence caused either by the resonant driving
or by a quench on top of the driving is an additional indicator for the observed phenomena. 
To obtain a comprehensive understanding of the microscopic properties of the strongly driven and interacting system, we focus on the  
few-body dynamics in small lattices (specifically, four bosons in a triple well setup). However, we provide strong evidence that our findings
apply equally to larger lattice systems and particle numbers.
All calculations to solve the underlying many-body Schr\"{o}dinger equation are performed by employing the  
Multi-Configuration Time-Dependent Hartree method for Bosons (MCTDHB) \cite{Alon,Alon1}, which is especially designed to treat the  
out-of-equilibrium quantum dynamics of interacting bosons under time-dependent modulations.

This work is organized as follows. In Sec.II we explain our
setup and introduce the multi-band expansion and the basic observables that we shall use in order to interpret the 
dynamics. Sec.III presents the effects resulting from an   
interaction quench of a driven triple well for filling factors larger than unity. Sec.IV. presents 
the dynamics for filling factors smaller than unity. We
summarize our findings and give an outlook in Sec.V. In Appendix A the non-equilibrium dynamics 
induced by a driven harmonic oscillator and 
simultaneously interaction quenched bosonic cloud is briefly outlined. 
Appendix B briefly comments on the resonant response of the driven lattice and finally Appendix C describes our computational method.

\section{Setup and analysis tools}

In the present section we shall briefly report on our theoretical framework. Firstly, we introduce  
the protocol of the driven optical lattice and the many-body Hamiltonian. Secondly, the wavefunction representation in terms 
of a multiband expansion and some basic observables for the understanding of the inter- and intra-well modes  
of the dynamics, are introduced.

\subsection{Setup and Hamiltonian}

To model a lattice vibration, with amplitude $\delta$
and angular frequency $\omega_{D}=2\pi f_{\rm{D}}$, a spatio-temporal sinusoidal modulation is used to
generate a lattice potential of the form
\begin{equation}
\label{eq:1}{V_{br}}({x};t) = {V_0}{\sin ^2}[k_x(1+\delta
\sin(\omega_{D}t))x],
\end{equation}
with lattice depth ${V_0}$ and 
wave-vector $k_x= \frac{\pi }{l}$, where $l$ denotes the
distance between successive potential minima.  
Such a potential can be realized e.g. via acousto-optical modulators \cite{Parker}, which induce  
a frequency difference among counterpropagating laser beams. 
The Hamiltonian of $N$ identical bosons of 
mass $M$ following an interaction quench protocol upon the driven one-dimensional lattice reads 
\begin{equation}
\label{eq:2}H(x;t) = \sum\limits_{i = 1}^N { \frac{{{p_i ^2}}}{{2M}}}
 + {V_{br}}({x_i};t) +
g_{1D}^{(f)}\sum\limits_{i < j} {\delta({x_i} - {x_j})},
\end{equation}
where $g_{1D}^{(f)}=\delta{g}+g_{1D}^{(in)}$, with $g_{1D}^{(in)}$, $g_{1D}^{(f)}$ being the initial and final interaction
strengths respectively and $\delta{g}$ denotes the corresponding perturbation. The short-range interaction
potential between particles located at positions ${x_i}$, 
is modeled by a Dirac delta-function. The
interaction is well described by s-wave scattering and the effective
1D coupling strength \cite{Olshanii} becomes ${g_{1D}} =
\frac{{2{\hbar ^2}{a_0}}}{{Ma_ \bot ^2}}{\left( {1 - \frac{{\left|
{\zeta (1/2)} \right|{a_0}}}{{\sqrt 2 {a_ \bot }}}} \right)^{ -
1}}$. The transversal length scale is ${a_ \bot } = \sqrt
{\frac{\hbar }{{M{\omega _ \bot }}}}$, with ${{\omega _ \bot }}$ the
frequency of the confinement, while ${a_0}$ denotes the 3D s-wave
scattering length. The interaction strength can be
tuned either via ${a_0}$ with the aid of Feshbach resonances
\cite{Kohler,Chin}, or via the transversal confinement frequency
${\omega _ \bot }$ \cite{Kim,Giannakeas,Giannakeas1}.

In the following, for reasons of universality  
the Hamiltonian (2) is rescaled in units of the recoil
energy ${E_{\rm{R}}} = \frac{{{\hbar ^2k_x^2}}}{{2M}}$. 
Then, the corresponding length, time and frequency scales are given in units of
${k_x^{ - 1}}$, $\omega_{\rm{R}}^{-1}=\hbar E_{\rm{R}}^{ - 1}$ and $\omega_{\rm{R}}$, respectively. For our
simulations we have used a sufficiently large lattice depth of the
order of $V_{0}=10.0E_{R}$, such that each well
includes three localized single-particle Wannier states. The
confinement of the bosons in the $m$-well system is imposed by the
use of hard-wall boundary conditions at the appropriate position
$x_{\sigma} = \pm \frac{{m\pi }}{{2k_x}}$. Finally, for computational convenience we shall set 
$\hbar = M = k_x = 1$ and therefore all quantities below are 
given in dimensionless units. 

\subsection{Wavefunction representation and basic observables}

To understand the microscopic properties and analyze the dynamics, the notion of   
non-interacting multiband Wannier number states is employed. The presently used lattice potential 
is deep enough for the Wannier states between different wells to have a very small overlap for
not too high energetic excitation. In the case of a periodically driven potential the above description can still be valid 
if the driving amplitude is small enough in comparison to the
lattice constant $l$, i.e. $\delta\ll l$ such that each localized Wannier
function is assigned to a certain well and the respective
band-mixing is fairly small. For $\delta \gg l$ the use of 
a time-dependent Wannier basis is more adequate.
Summarizing, for a system with $N$ bosons, $m$-wells and $j$ localized single
particle states \cite{Mistakidis,Mistakidis1} the expansion of the many-body 
bosonic wavefunction reads
\begin{equation}
\label{eq:3}\left| \Psi  \right\rangle  = \sum\limits_{\{N_{i}\},\{I_i\}}
{{C_{\{N_i\};\{I_i\}}}{{\left| {{N_1^{(I_1)}},{N_2^{(I_2)}},...,{N_m^{(I_m)}}} \right\rangle
}}},
\end{equation}
where ${{\left| {{N_1^{(I_1)}},{N_2^{(I_2)}},...,{N_m}^{(I_m)}} \right\rangle}}$ is the multiband Wannier number state, the element 
${{N_i}^{(I_i)}}=\ket{n_i^{(1)}}\otimes\ket{n_i^{(2)}}\otimes....\otimes\ket{n_i^{(j)}}$
denotes the number of bosons being localized in the
$i$-th well, and $I_i$ indexes the corresponding energetic excitation order. In particular, 
$\ket{n_i^{(k)}}$  
refers to the number of bosons which reside at
the $i$-th well and $k$-th band, satisfying the closed subspace constraint
$\sum_{i=1}^m\sum_{k=1}^j n_i^{(k)} = N$.
For instance, in a setup with $N=4$ bosons confined in a triple well
i.e. $m=3$, which includes $k=3$ single particle states, the state 
$\ket{1^{(0)} \otimes 1^{(1)},1^{(0)},1^{(0)}}$
indicates that in every well 
one boson occupies the zeroth excited band, but in the left well there is one extra boson  
localized in the first excited band. 
For this setup it is also important to notice that one can realize
four different energetic classes of number states, namely the quadruple mode $\{ {\left| {4^{(I_1)},0^{(I_2)},0^{(I_3)}}
\right\rangle}+\circlearrowright\}$ (Q),
the triple mode $\{ {\left| {3^{(I_1)},1^{(I_2)},0^{(I_3)}} \right\rangle}+\circlearrowright\}$ (T), the
double pair mode $\{ {\left| {2^{(I_1)},2^{(I_2)},0^{(I_3)}} \right\rangle}+\circlearrowright\}$ (DP), 
and the single pair mode $\{ {\left| {2^{(I_1)},1^{(I_2)},1^{(I_3)}}\right\rangle}+\circlearrowright\}$ (SP), where $\circlearrowright$ stands for all corresponding permutations.
It is important to note that, for later convenience, we consider only the corresponding subclass with isoenergetic 
states and not all members which would also include energetically unequal number states, e.g. for the single pair mode 
$\{ {\left| {2^{(I_1)},1^{(I_2)},1^{(I_3)}}
\right\rangle}, {\left| {1^{(I_1)},2^{(I_2)},1^{(I_3)}} \right\rangle}, {\left| {1^{(I_1)},1^{(I_2)},2^{(I_3)}}
\right\rangle}\}$.
Also, in the present consideration for a given set of excitation indices $\textbf{I}=(I_1,I_2,I_3)$, the above-mentioned class of number states we are focusing on 
have similar on-site energies and will contribute significantly to
the same eigenstates. 
Indexing each such class by $\alpha$, we adopt
the more compact notation ${\left|{q} \right\rangle
_{\alpha;\textbf{I}}}$ for the characterization of the eigenstates in terms
of number states, where the index $q$ refers to the spatial
occupation. For instance ${\{\left|{q} \right\rangle _{3;\textbf{I}}}\}$ with $\textbf{I}=(1,0,0)$ 
represent the eigenstates which are dominated by the set of the
triple pair states $\{ {\left| {3^{(1)},1^{(0)},0^{(0)}} \right\rangle}$, ${\left| {0^{(0)},3^{(1)},1^{(0)}}
\right\rangle}$, ${\left| {1^{(0)},0^{(0)},3^{(1)}} \right\rangle}$, ${\left| {1^{(0)},3^{(1)},0^{(0)}}
\right\rangle}$, ${\left| {0^{(0)},1^{(0)},3^{(1)}} \right\rangle}$, ${\left| {3^{(1)},0^{(0)},1^{(0)}}
\right\rangle}\}$, and the index $q$ runs from 1 to 6.

Below, a few basic observables which refer to the inter- and intra-well generated modes are introduced and   
their expansion in terms of the multiband number state basis is given.  
Note that from here on we shall denote by $\left|
{\Psi (0)} \right\rangle  = \sum\limits_{q ;\alpha ;\textbf{I}} {C_{\alpha ;\textbf{I}}
^q{{\left| q \right\rangle }_{\alpha ;\textbf{I}}}}$ the
initial wavefunction in terms of the eigenstates ${{{\left| q
\right\rangle }_{\alpha ;\textbf{I}}}}$ of the final Hamiltonian. 
A time resolved measure for the 
impact of the external driving on the system is provided via the fidelity $F_{\{\lambda_i\}}(t)={\left| {\left\langle {\Psi (0)} \right|\left. {\Psi_{\{\lambda_i\}}
(t)} \right\rangle } \right|^2}$, being the overlap between the 
time evolved and the initial (ground) state.
Note the dependence of the fidelity on the set of parameters $\{\lambda_{i}\}$, e.g.   
the driving frequency $\omega_{D}$, the interaction strength $g$, the 
particle number $N$ etc. The expansion of the fidelity reads
\begin{equation}
\begin{split}
\label{eq:9}F_{\{\lambda_i\}}(t) = \sum\limits_{q_{1};\alpha;\textbf{I}}
{{\left| {{C_{\alpha;\textbf{I}}^{q_{1}}}} \right|}^4} +
\sum\limits_{q_{1},q_{2};\alpha ,\beta;\textbf{I}} {{\left|
{{C_{\alpha;\textbf{I}}^{q_{1}}}} \right|}^2}\\\times{{\left|
{{C_{\beta;\textbf{I}}^{q_{2}}}} \right|}^2}\cos ({\epsilon
_{\alpha;\textbf{I}}^{q_{1}}} - {\epsilon _{\beta;\textbf{I}}^{q_{2}}})t.
\end{split}
\end{equation}
The second term on the right-hand-side of the above expression contains the energy difference between two distinct number states and 
therefore offers to be a measure of the tunneling process. 
The indices $\alpha$, $\beta $ indicate a
particular number state group \cite{Mistakidis}, ${q _i}$ is the
intrinsic index within each group, $\textbf{I}$ corresponds to the respective
energetical level and $\epsilon$ refers to the corresponding on-site
energy of a particular number state and energetical level. 

For the investigation of the intra-well dynamics it is appropriate to employ a local density analysis. 
To measure the instantaneous spreading of the cloud in the $i$-th well we
define the operator of the second moment $\sigma _i^2(t) =
\left\langle {{{\Psi|\left( {x - R_{CM}^{(i)}} \right)}^2}}
|\Psi\right\rangle $ \cite{Ronzheimer}.
Here $R_{CM}^{(i)} =\int_{d_i}^{d'_{i}}dx \left( {x - x_0^{(i)}}
\right){\rho _i}(x)/\int_{d_i}^{d'_{i}}dx{\rho _i}(x)$ refers to 
the coordinate of the center-of-mass \cite{Klaiman1,Klaiman2},   
${x_0^{(i)}}$ denotes the central point of the
$i$-th well under investigation, ${d_i}$, 
${d'_i}$ correspond to the instantaneous limits of the wells, whereas ${\rho _{i}}(x)$ is the respective single
particle density. Then, the expansion of the second moment for the middle well in terms of 
the eigenstates of the final Hamiltonian reads 
\begin{equation}
\begin{split}
\label{eq:10}\begin{array}{l}
\sigma _M^2(t) = \sum\limits_{\alpha ;q_{1};\textbf{I}} {{{\left| {C_{a;\textbf{I}}^{q_{1}}} \right|}^2}{}_{\alpha ;\textbf{I}}\left\langle q_{1} \right|} {\left( {x - R_{CM}^{(i)}} \right)^2}{\left| q_{1} \right\rangle _{\alpha ;\textbf{I}}} \\
 ~~~~~~~~+ 2\sum\limits_{q_{1} \ne q_{2}} {{\mathop{\rm Re}\nolimits} \left( {C_{\beta ;\textbf{I}}^{*q_{1}}C_{\alpha ;\textbf{I}}^{q_{2}}} \right){}_{\beta ;\textbf{I}}\left\langle q_{1} \right|} {\left( {x - R_{CM}^{(i)}} \right)^2}{\left| q_{2} \right\rangle _{\alpha ;\textbf{I}}}\\~~~~~~~~~~~~~~~~~~~~~~~~~~~~~\times\cos \left( {\omega _{\beta ;\textbf{I}}^{q_{1}} - \omega _{\alpha ;\textbf{I}}^{q_{2}}} \right)t.
\end{array}
\end{split}
\end{equation}

Finally, as a measure of the dipole motion the intra-well asymmetry $\Delta
{\rho _a}(t) = {\rho _{a,1}}(t) - {\rho _{a,2}}(t)$ is introduced.
Here, a particular well $a$ (in a triple well $a = L,M,R$ stands for the left, middle and
right well respectively) is divided from the center point into two equal sections
with ${\rho _{a,1}}(t)$ and ${\rho _{a,2}}(t)$ being the respective
integrated densities of the left and right parts during the
evolution.  
The expectation value of the asymmetry operator is expressed as 
\begin{equation}
\begin{split}
\label{eq:13}\begin{array}{l}\left\langle \Psi  \right|\Delta
\rho(t) \left| \Psi  \right\rangle  =
\sum\limits_{q_{1};\alpha;\textbf{I}}  {{{\left|
{{C_{\alpha;\textbf{I}}^{q_{1}}}} \right|}^2}{}_{\textbf{I};\alpha}\left\langle
q_{1} \right|} \Delta\rho \left| q_{1}
\right\rangle_{\alpha;\textbf{I}} \\
~~~~~~~~+ 2\sum\limits_{q_{1}  \ne q_{2} } {{\mathop{\rm
Re}\nolimits} \left( {C_{\alpha;\textbf{I}
}^{*q_{1}}{C_{\beta;\textbf{I}}^{q_{2}}}} \right){}_{\textbf{I};\alpha}\left\langle q_{1}  \right|}
\Delta \rho \left| q_{2}
\right\rangle_{\beta;\textbf{I}}\\~~~~~~~~~~~~~~~~~~~~~~~~~~~~~~\times \cos
\left[ {\left( {{\omega _{\alpha;\textbf{I}}^{q_{1}} } - {\omega
_{\beta;\textbf{I}}^{q_{2}}}} \right)t} \right].
\end{array} 
\end{split}
\end{equation}

\subsection{First order coherence}

The spectral representation of the reduced one-body density matrix \cite{Titulaer,Naraschewski,Sakmann4} reads 
\begin{equation}
\label{eq:4} \rho_1 (x,x';t) = \sum\limits_{\alpha=1}^{M}
{{n_{\alpha}}(t){\varphi _{\alpha}}(x,t)} \varphi _{\alpha}^*(x',t), 
\end{equation}
where ${\varphi _\alpha}(x,t)$ are the so-called natural orbitals and $M$  
corresponds to the considered number of orbitals. The population
eigenvalues $n_{\alpha}(t) \in [0,1]$ characterize the fragmentation of the
system \cite{Spekkens,Klaiman,Mueller,Penrose}: For only one
macroscopically occupied orbital the system is said to be condensed, 
otherwise it is fragmented.

To quantify the degree of first order coherence during the dynamics, the normalized spatial first order correlation 
function $g^{(1)}(x,x';t)$ is defined  
\begin{equation}
\label{eq:4} g^{(1)}(x,x';t)=\frac{\rho_{1}(x,x';t)}{\sqrt{\rho_1(x;t)\rho_1(x';t)}}.
\end{equation}
It is known that for $|g^{(1)}(x,x';t)|^2<1$ the corresponding visibility of interference
fringes in an interference experiment is less than $100\%$ and this case is referred to as loss 
of coherence. On the contrary, when $|g^{(1)}(x,x';t)|^2=1$ the fringe visibility of the interference
pattern is maximal and is referred to as full coherence. The above quantity depends 
strongly on the various parameters of the Hamiltonian, and an investigation of the aforementioned dependence will be 
given in Sec. IV.

\section{Interaction quench dynamics on a driven lattice for filling factor $\nu>1$}

To analyze the dynamics of our system, it is instructive first to  
comment on the relation between the ground state and its dominant interaction dependent spatial 
configuration employing the multiband expansion. Let us consider a  
setup with four bosons in a triple well (which is our workhorse).   
Within the weak interaction regime $0<g<0.1$ the dominant spatial configuration of the 
system is $\ket{1^{(0)},2^{(0)},1^{(0)}}$, 
while states of double-pair, e.g. $\ket{2^{(0)},2^{(0)},0^{(0)}}$, and triple-pair 
occupancy, e.g. $\ket{1^{(0)},3^{(0)},0^{(0)}}$, possess a small contribution. In the intermediate interaction regime $0.1<g<1.0$  
the system is described by a superposition of lowest-band states 
which are predominantly of single-pair occupancy, e.g. $\ket{1^{(0)},2^{(0)},1^{(0)}}$, $\ket{2^{(0)},1^{(0)},1^{(0)}}$, and 
double-pair occupancy, e.g. $\ket{2^{(0)},2^{(0)},0^{(0)}}$, while 
energetically higher states to the first excited-band start to be occupied. 
For further increasing repulsion, e.g. $1.0<g<5.0$, the excited states gain more population 
and the corresponding ground state configuration is characterized by an admixture of 
ground- (predominantly of single pair occupancy) and excited-band (to the first and even to the second band) states.

In the following, we shall investigate the effect of an interaction 
quench upon a periodically driven finite lattice. Note that we consider interaction quenches 
imposed at $t=0$ or after a short transient time. The resulting dynamics is qualitatively   
the same. We shall refer, for brevity,    
to the effect of an interaction quench performed at $t=0$, i.e. when also the periodic driving starts.    
To be more specific, below we shall 
firstly explore the effect of an interaction quench for various driving frequencies and compare the induced dynamics with an unquenched 
system. Subsequently, the dynamics for a fixed driving frequency while varying the quench amplitude is investigated. We remark   
that in each case we consider quench amplitudes for which the induced above-barrier transport is suppressed. 
\begin{figure*}[ht]
        \centering
           \includegraphics[width=0.80\textwidth]{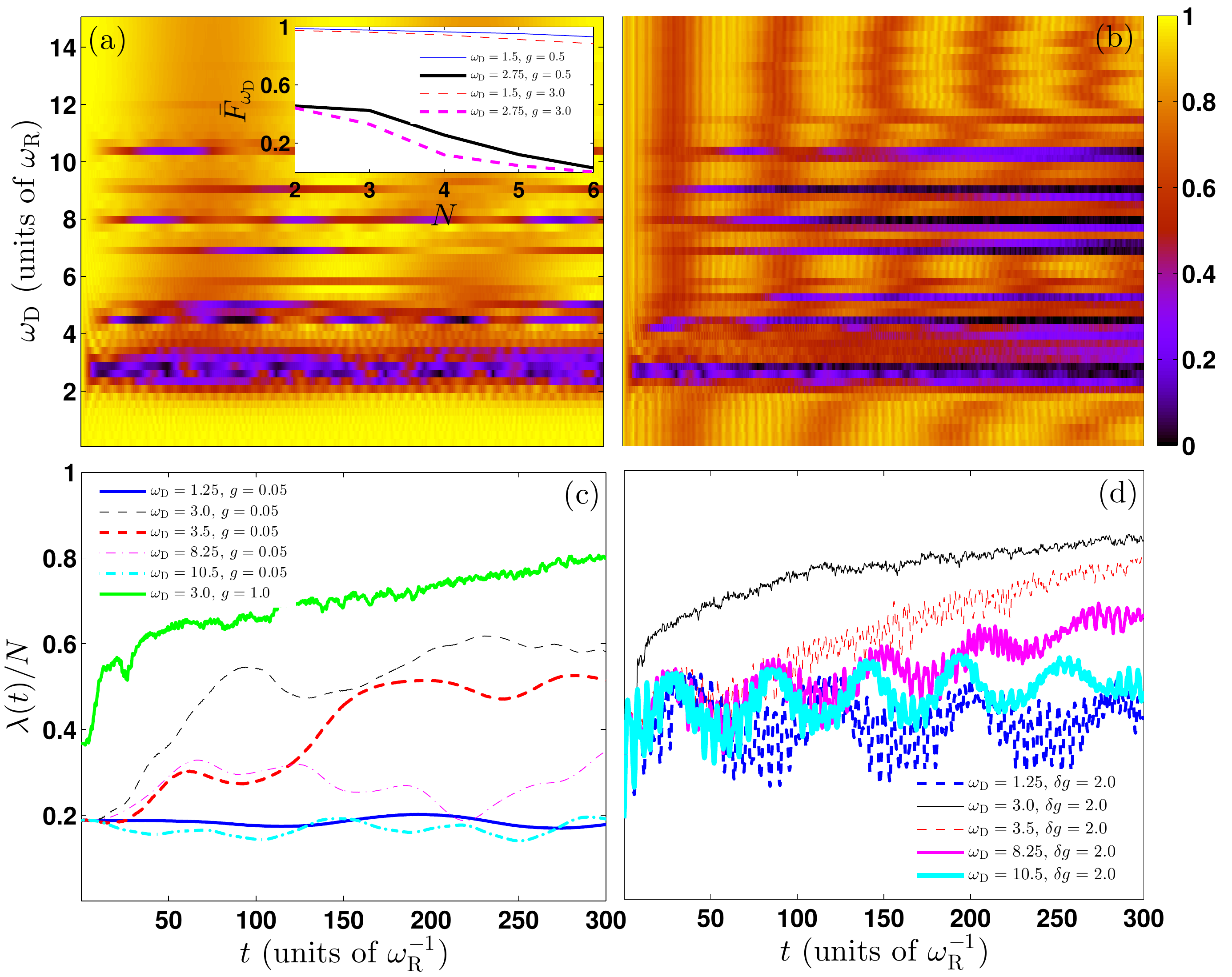}
                \caption{(a) Time evolution of the fidelity $F_{\{\omega_{D}\}}(t)$ as a function of the driving frequency $\omega_{D}$.
                The driving amplitude is $\delta=0.03$ and the initial state corresponds to the ground state of four
                weakly interacting bosons with $g=0.05$ confined in a triple well. Inset: Mean response $\bar{F}_{\{\omega_{\rm{D}}\}}$ at $\omega_{\rm{D}}=0.75$ and at $\omega_{\rm{D}}=2.75$  
                for different interparticle repulsion $g=0.5$ and $g=3.0$ as a function of the particle number $N$ (see legend). (b) Same as (a) but for a fixed interaction quench, with amplitude $\delta g=2.0$, on top of the 
                driven triple well. (c) Deviation from unity of the first natural occupation number, i.e.  
                $\lambda(t)=1-n_1(t)$ during the evolution for different driving frequencies $\omega_{\rm{D}}$ (see legend). The effect of a stronger interparticle repulsion 
                for $g=1.0$ at $\omega_{\rm{D}}=3.0$ in the 
                fragmentation process is also shown. (d) The same as (c) but for a fixed interaction quench, $\delta g=2.0$, upon the driving.}
\end{figure*}

\subsection{Case I: Interaction quench dynamics for different driving frequencies}

We shall explore the effect of an interaction quench on top 
of a periodically driven triple well potential 
with four bosons in the weak interaction regime  
($g= 0.05$), where the dominant spatial configuration of the ground state    
corresponds to states of single-pair occupancy e.g. $\ket{1^{(0)},2^{(0)},1^{(0)}}$.
To demonstrate the difference between the dynamics of the quenched and the unquenched bosonic ensemble let us firstly  
investigate the response of an explicitly driven system, i.e. with $\delta g=0$. Figure 1(a) shows $F_{\{\omega_{\rm{D}}\}}(t)$ (see also Sec. II.B) 
with varying $\omega_{D}$. It is observed that for $0<\omega_{D}<1.5$ (nearly adiabatic driving) or very intense 
driving $\omega_{D}>12.0$ the system remains essentially unperturbed. 
In between, an interesting stripe pattern occurs. To be self-contained, in the following, let us classify the frequency intervals 
\begin{equation}
\label{eq:4}\Delta\omega_{\rm{D}_1}\equiv[2.0,6.0]~~and~~\Delta \omega_{\rm{D}_2}\equiv[7.0,11.0],  
\end{equation}
where the time evolved state of the periodically driven system deviates significantly from the initial (ground) state. Indeed, for $\omega_{\rm{D}}\in\Delta\omega_{\rm{D}_1}\equiv[2.0,6.0]$ 
the minimal overlap during the dynamics drops down to 0.1, whereas for $\omega_{\rm{D}}\in\Delta \omega_{\rm{D}_2}\equiv[7.0,11.0]$  
the system maximally departs from the initial state by a percentage of the order of $30\%$. 
To probe the effect of the interactions and of the driving frequency on the overall dynamics,  
the inset of Figure 1(a) illustrates $\bar{F}_{\{\omega_{\rm{D}}\}}=\int_{0}^{T}dtF_{\{\omega_{\rm{D}}\}}(t)/T$, ($T$ denotes the considered evolution 
time) at $\omega_{\rm{D}}=1.5$ and at $\omega_{\rm{D}}=2.75\in\Delta\omega_{\rm{D}_1}$ for different initial interactions and particle number.  
Focusing on the same driving frequency $\omega_{\rm{D}}$ and a large interparticle interaction we observe 
that the mean response of the system decreases as a function of the particle number and therefore the system can be driven more efficiently 
out-of-equilibrium. The same observation holds for a fixed interaction strength and particle number but a driving frequency below and in the region $\Delta\omega_{\rm{D}_1}$,   
e.g. for $N=4$, $g=3$, $\bar{F}_{\{\omega_{\rm{D}}=1.5\}}=0.9405$, while $\bar{F}_{\{\omega_{\rm{D}}=2.75\}}=0.1202$.   
Let us now inspect how an interaction quench distorts the fidelity evolution.   
Figure 1(b) shows $F_{\{\omega_{\rm{D}},\delta g\}}(t)$ for $\delta{g}=2.0$ (performed at $t=0$, i.e. simultaneously with the driving) with varying $\omega_{\rm{D}}$. It is observed that the combination of driving and interaction quench   
brings the system significantly out-of-equilibrium for 
every driving frequency. 
To understand the effect of the quench on the system let us compare Figure 1(b) with    
Figure 1(a) for the fidelity evolution of the driven but unquenched system.  
Indeed, an interaction quench introduces more energy into the system and as a consequence the final evolving state deviates 
significantly from the initial one even 
in the region of adiabatic driving, e.g. $\omega_{\rm{D}}=0.5$ or high    
frequency driving, e.g. $\omega_{\rm{D}}=14.0$, as seen in Figure 1(b). For instance,  
$\bar{F}_{\{\omega_{\rm{D}}=1.0,\delta g=0\}}=0.98$ and $\bar{F}_{\{\omega_{\rm{D}}=1.0,\delta g=2.0\}}=0.81$, while $\bar{F}_{\{\omega_{\rm{D}}=14.0,\delta g=0\}}=0.92$ 
and $\bar{F}_{\{\omega_{\rm{D}}=14.0,\delta g=2.0\}}=0.78$. Finally, as 
an estimate we report that according to our simulations the deviation of $\bar{F}$ between the unquenched and the quenched system ranges from $12\%$ to $70\%$.  

To analyze the role of dynamical fragmentation \cite{Mueller,Sakmann5} (see Eq.(7)), 
Figure 1(c) shows the deviation from unity, $\lambda(t)=1-n_1(t)$, during the evolution of the first natural
population for different driving frequencies $\omega_{\rm{D}}$ and no quench.  
Note here that even $\lambda(0)\neq 0$, i.e. as a result 
of the finite repulsion the initial state 
possesses a small degree of fragmentation.
As shown $\lambda(t)$, is always significantly above zero,  
confirming the fragmentation process. 
Focusing on different $\omega_{\rm{D}}$'s   
we note that the temporal average of the fragmentation, i.e. $\bar{\lambda}= \int dt\lambda(t)/T$, increases if $\omega_{\rm{D}}\in\Delta\omega_{\rm{D}_1}\cup\Delta\omega_{\rm{D}_2}$, while for 
the regions where $F_{\{\omega_{\rm{D}}\}}\simeq1$ it reduces but never tends to a perfectly 
condensed state. Note also that for $\omega_{\rm{D}} \not\in \Delta\omega_{\rm{D}_1}\cup\Delta\omega_{\rm{D}_2}$,   
$\lambda(t)$ possesses small amplitude oscillations, whereas for $\omega_{\rm{D}}$ $\in \Delta\omega_{\rm{D}_1}\cup\Delta\omega_{\rm{D}_2}$ the external driving 
introduces large amplitude variations in $\lambda(t)$. 
As expected the interparticle repulsion supports the fragmentation process (see $\lambda(t)$ for $\omega_{\rm{D}}=3.0$, $g=1.0$ and 
$\delta g=0.0$ in Figure 1(c)). The effect of an interaction quench on the fragmentation process 
is shown in Figure 1(d) employing $\lambda(t)$, for $\delta g=2.0$ and the same driving frequencies as in Figure 1(c). 
A tendency for a higher fragmented state for every $\omega_{\rm{D}}$ at least for certain time periods is manifest.  
Comparing $\lambda(t)$ for $\omega_{\rm{D}}$ below $\Delta \omega_{\rm{D}_1}$, with the unquenched case, we observe that the interaction quench 
introduces large amplitude variations, while for $\omega_{\rm{D}}\in\Delta \omega_{\rm{D}_1}\cup\Delta \omega_{\rm{D}_2}$ $\lambda(t)$ shows a monotonic increase towards a fully fragmented state. 
Thus, in conclusion, the fragmentation process under an interaction quench is enhanced, 
which is attributed to the consequent raise of the interparticle repulsion. 

To identify the effect of an interaction quench on the one-body level, Figure 2 compares $\rho_1(x,t)$ without and with an interaction quench on top of the 
periodically driven triple well for 
$\omega_{\rm{D}}=0.75$ and amplitude $\delta=0.03$. Without quench, the one-body density (see Figure 2(a)) shows a weak response, a local dipole mode in the outer wells and a local breathing mode 
(hardly visible in Figure 2(a) due to weak driving) in the central well occur  
due to the combination of the parity of the lattice (odd number of sites) and 
the driving scheme. The dynamics in the central well shows a compression and decompression, while the outer wells are shaken  
(for a lattice with an even number of sites the generated intra-well mode will solely be a local dipole mode). 
As can be seen, by performing a quench (see Figure 2(b)) with $\delta g=2.0$, the breathing-like mode in the central well is enhanced, while in the outer wells the cloud exhibits admixtures  
of excitations consisting of a dipole and a breathing component. Focusing on the dynamics of the left well it 
is obvious that the atomic cloud oscillates 
inside the well with a varying 
amplitude, i.e. it performs an oscillation 
with a simultaneous compression and decompression.  
Finally, the inter-well tunneling mode which is  
manifested as a direct population transport from the middle to the outer wells and accompanies the 
whole process is amplified. To illustrate explicitly the evolution of the atomic cloud 
in each well we follow the $\rho_1(x,t)=0.25$ of the local density, shown as the thick white line 
on top of the density. It is observed that in the central well the cloud compresses and decompresses during the 
evolution, while in the outer wells the cloud oscillates changing also its width (in Appendix A,  
this mode is generated in a harmonic trap for a deeper understanding). 
\begin{figure}[ht]
        \centering
           \includegraphics[width=0.50\textwidth]{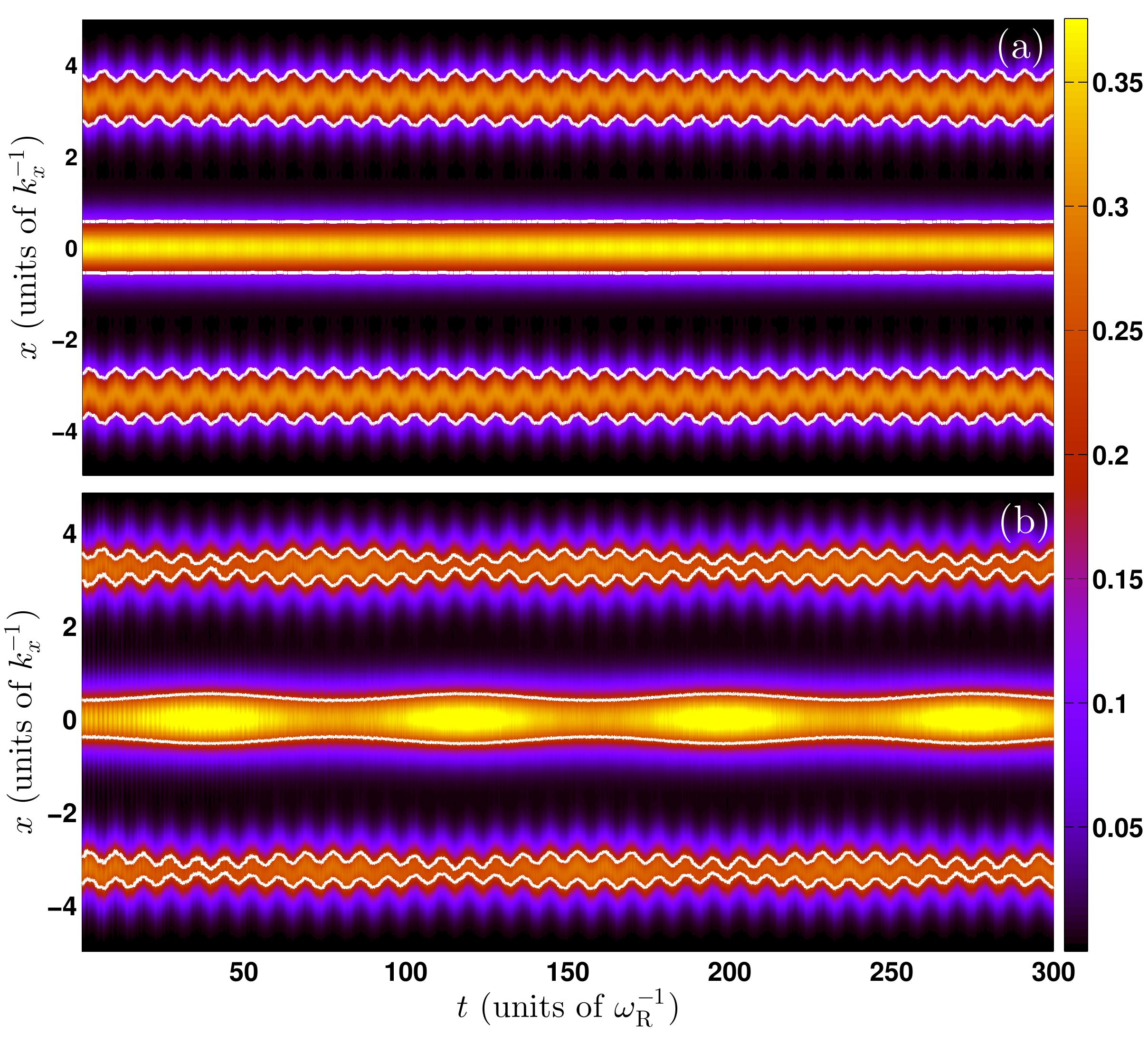}
                \caption{Time evolution of the one-body density $\rho_1(x,t)$ caused by a periodically driven triple well with (a) $\omega_{\rm{D}}=0.75$ and (b) a simultaneous 
                interaction quench with amplitude $\delta g=2.0$. White contours, at $\rho_1(x,t)=0.25$, are plotted on top in order to facilitate a 
                comparison of the atomic motion between the unquenched (a) and the quenched system (b).    
                The driving amplitude is fixed to the value $\delta=0.03$ and the initial state corresponds to the ground state of four
                weakly interacting bosons with $g=0.05$.}
\end{figure}

\begin{figure}[ht]
        \centering
           \includegraphics[width=0.40\textwidth]{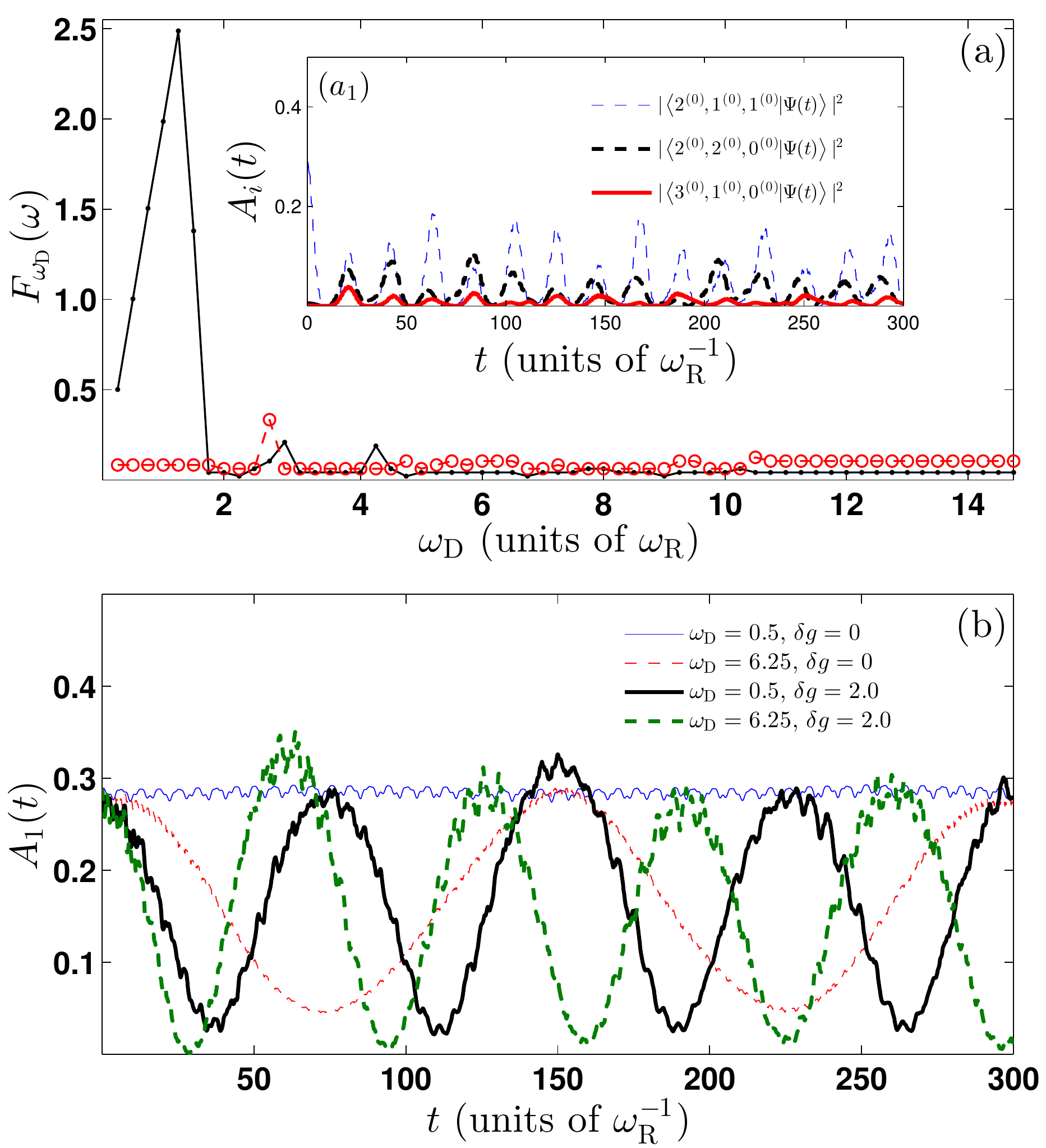}
                \caption{(a) Spectrum of the fidelity $F_{\{\omega_{\rm{D}}\}}(\omega)$ as a function of the driving frequency $\omega_{\rm{D}}$. The black dots 
                correspond to $F_{\{\omega_{\rm{D}}, \delta g=0.0\}}(\omega)$, i.e. to the 
                unquenched system, while 
                the red empty circles refer to $F_{\{\omega_{\rm{D}}, \delta g=2.0\}}(\omega)$ 
                , i.e. to the case of a simultaneous interaction quench with amplitude $\delta g=2.0$ on top of the driving. Inset 
                ($a_1$): Tunneling probabilities $A_1(t)$, $A_2(t)$ and $A_3(t)$ (see main text and legend) at $\omega_{\rm{D}}=2.75$. (b) Comparison of the single-particle tunneling probabilities $A_1(t)$ in a 
                periodically driven triple well without and with a simultaneous interaction quench for various driving frequencies $\omega_{\rm{D}}$ (see legend).  
                The driving amplitude is fixed to the value $\delta=0.03$ and the initial state corresponds to the ground state of four
                weakly interacting bosons with $g=0.05$.}
\end{figure}

\begin{figure*}[ht]
        \centering
           \includegraphics[width=0.70\textwidth]{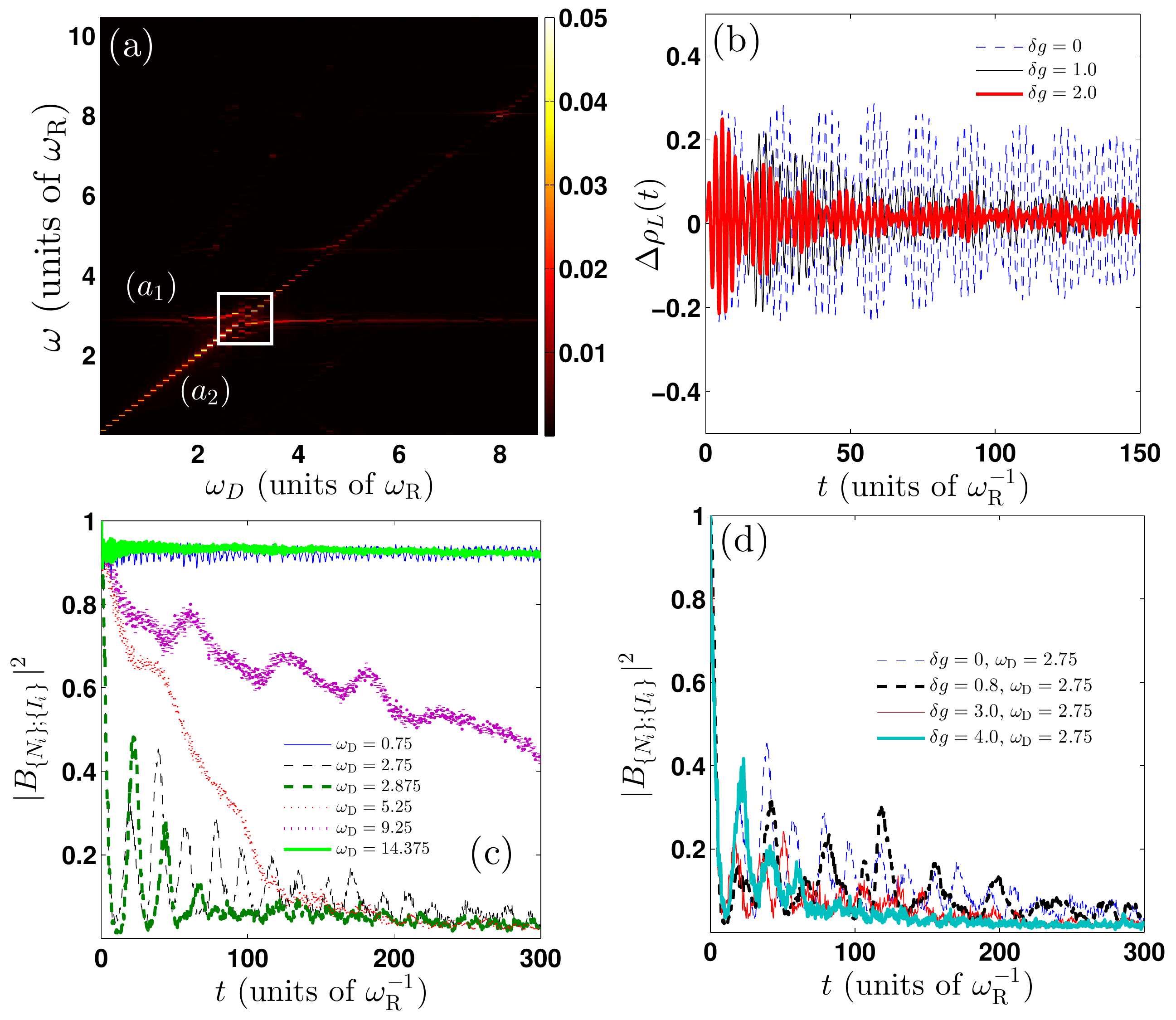}
                \caption{(a) Spectrum of the intra-well asymmetry for the left well $\Delta\rho_L(\omega)$ in a driven triple well, with respect to the driving frequency $\omega_{\rm{D}}$.   
                The white rectangle indicates the region of the resonance. (b) Intra-well asymmetry evolution $\Delta\rho_L(t)$ at resonance 
                ($\omega_{\rm{D}}=2.875$) employing different interaction quenches (see legend). (c) Excitation 
                probability $|B_{\{N_i\};\{I_i\}}|^2$ (see main text) during the evolution for different driving frequencies $\omega_{\rm{D}}$. (d) The same as (c) at $\omega_{\rm{D}}=2.75$ for different quenches 
                on the interparticle repulsion (see legend). The
                driving amplitude is fixed to the value $\delta=0.03$, while the initial state corresponds to the ground state of four
                weakly interacting bosons with $g=0.05$.}
\end{figure*}

\begin{figure}[ht]
        \centering
           \includegraphics[width=0.45\textwidth]{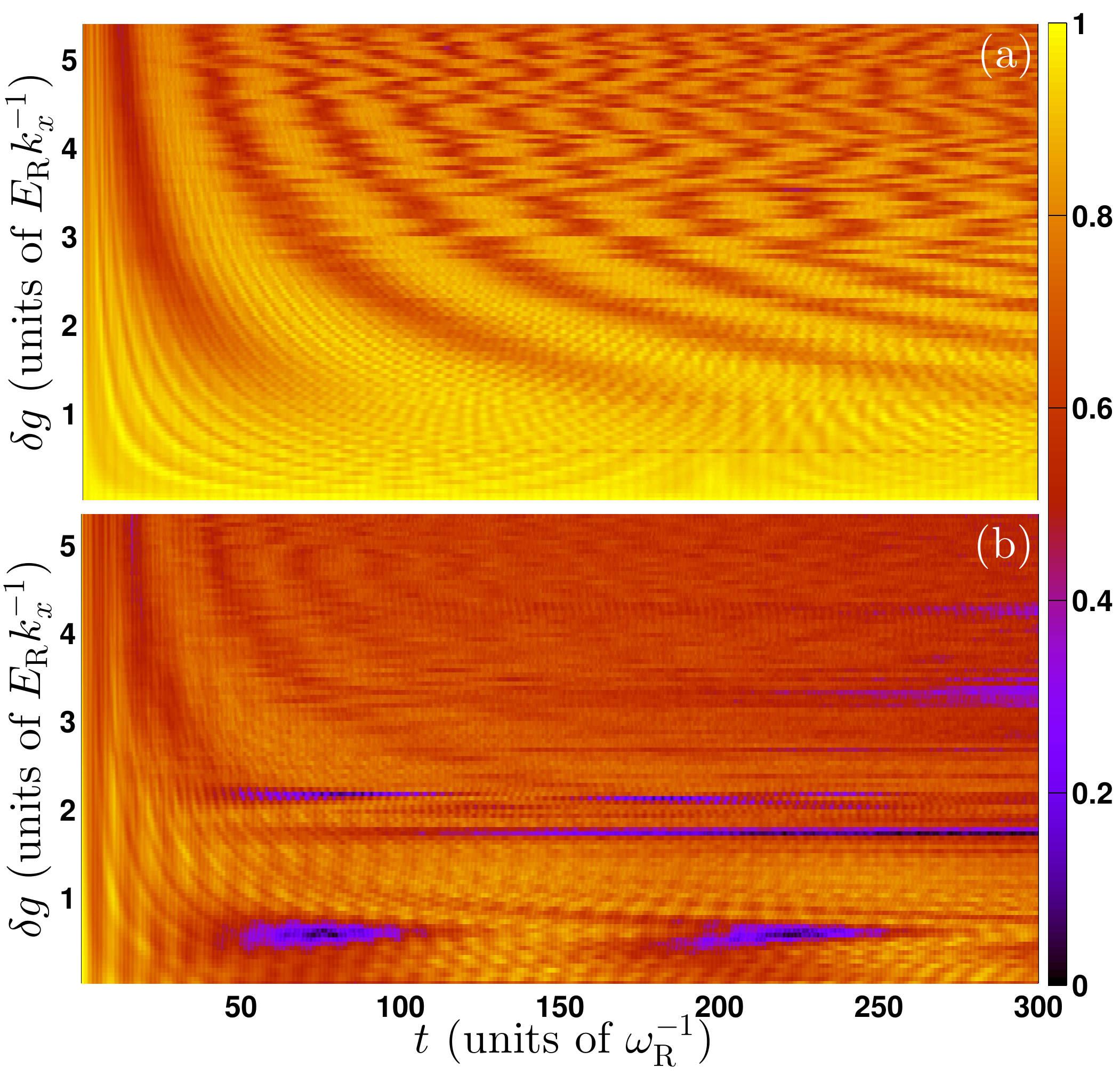}
                \caption{Time evolution of the fidelity $F_{\{\omega_{\rm{D}},\delta g\}}(t)$ in a periodically driven triple well with (a) $\omega_{\rm{D}}=0.75$ and (b) $\omega_{\rm{D}}=2.75$ as a function of the 
                quench amplitude. The
                driving amplitude is $\delta=0.03$, while the initial state corresponds to the ground state of four
                weakly interacting bosons with $g=0.05$.}
\end{figure}

\begin{figure*}[ht]
        \centering
           \includegraphics[width=0.50\textwidth]{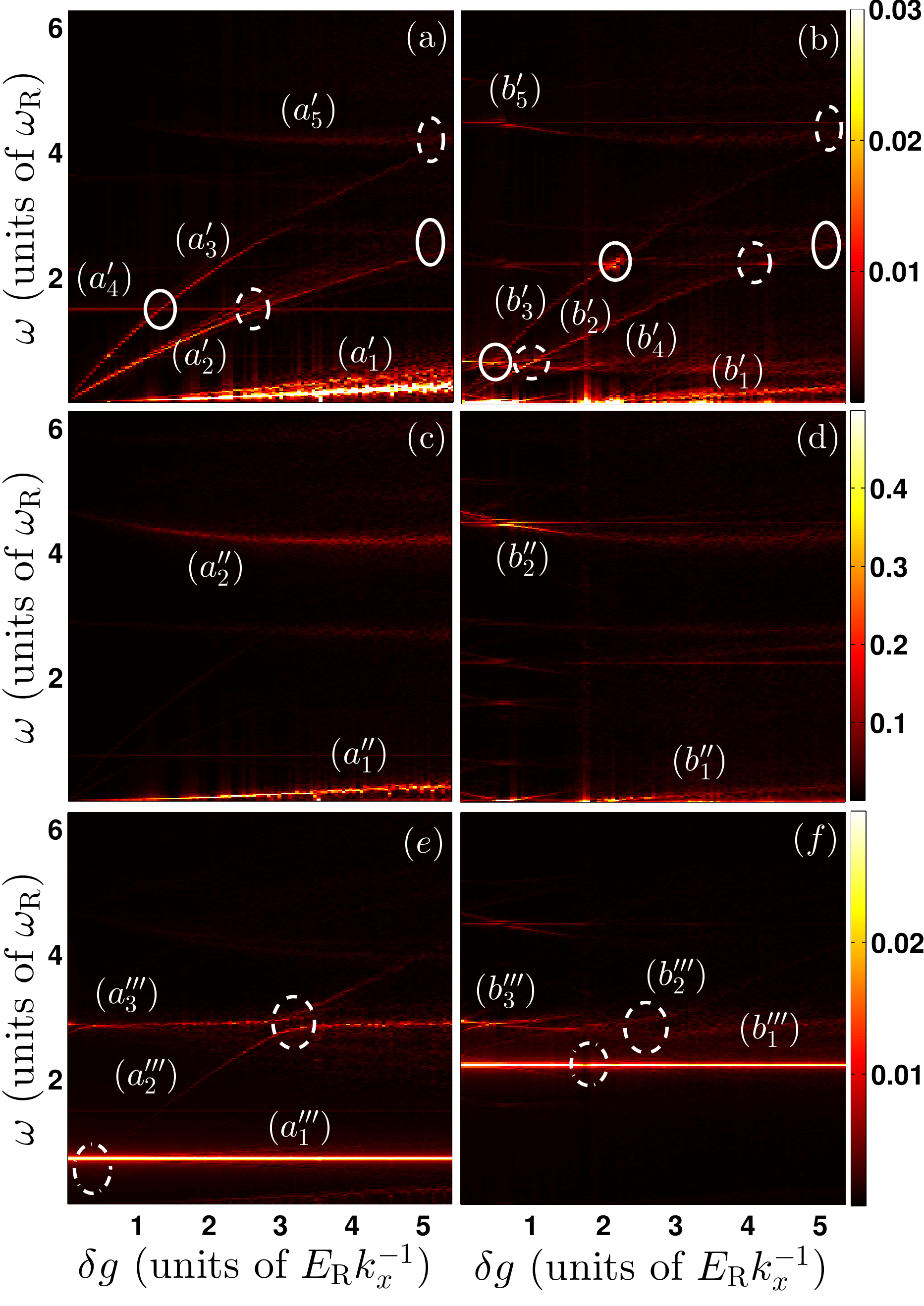}
                \caption{ As a function of the quench amplitude $\delta g$ are shown: fidelity spectrum $F_{\{\omega_{\rm{D}},\delta g\}}(\omega)$ for (a) $\omega_{\rm{D}}=0.75$ and (b) $\omega_{\rm{D}}=2.75$. 
                (c,d) Spectrum of the local-breathing mode ($\sigma_{M}^2(\omega)$) for the middle well of 
                the periodically driven triple well with (c) $\omega_{\rm{D}}=0.75$  
                and (d) $\omega_{\rm{D}}=2.75$. (e,f) Spectrum of the local-dipole mode ($\Delta\rho_L(\omega)$) for the left well of the   
                driven triple well with (e) $\omega_{\rm{D}}=0.75$ and (f) $\omega_{\rm{D}}=2.75$. The solid and dashed elipses indicate the positions of the resonances between the tunneling and the breathing or 
                dipole branches (see main text). The
                driving amplitude is $\delta=0.03$, while the initial state corresponds to the ground state of four
                weakly interacting bosons with $g=0.05$.}
\end{figure*}

To obtain a quantitative understanding of the inter-well tunneling dynamics, let us investigate the spectrum of the 
fidelity, i.e. ${F_{\{\omega_{\rm{D}}, \delta g\}}}(\omega ) = \frac{1}{\pi }\int {dt{F_{\{\omega_{\rm{D}}, \delta g\}}
}(t){e^{i\omega t}}}$ (see also Eq.(4)). Figure 3(a) shows both the tunneling spectrum of the unquenched (see black line) 
and the quenched system (see red line) with respect to the driving frequency. Indeed, employing Eq.(4) we obtain that for the unquenched system the dominant 
tunneling process for every $\omega_{\rm{D}}$ corresponds to tunneling within the 
SP mode (e.g. between the states $\ket{2^{(0)},1^{(0)},1^{(0)}}$ and $\ket{1^{(0)},2^{(0)},1^{(0)}}$). It is 
important here to note that for $\omega_{\rm{D}}\in\Delta\omega_{\rm{D}_1}$ additional tunneling modes  
from the SP to the DP mode (e.g. from $\ket{1^{(0)},2^{(0)},1^{(0)}}$ to $\ket{2^{(0)},2^{(0)},0^{(0)}}$) and from 
the SP to the T mode (e.g. from $\ket{1^{(0)},2^{(0)},1^{(0)}}$ to $\ket{3^{(0)},1^{(0)},0^{(0)}}$) can be generated.  
To ilustrate this fact we depict in the inset of Figure 3(a) the probabilities $A_1(t)=|\left\langle
2^{(0)},1^{(0)},1^{(0)}| \Psi(t) \right\rangle|^2$, $A_2(t)=|\left\langle
2^{(0)},2^{(0)},0^{(0)}| \Psi(t) \right\rangle|^2$ and $A_3(t)=|\left\langle
3^{(0)},1^{(0)},0^{(0)}| \Psi(t) \right\rangle|^2$ at $\omega_{\rm{D}}=2.75$. It is shown that $A_{2}(t)$ and $A_{3}(t)$ although suppressed in 
comparison to $A_1(t)$ possess significant populations. We remark here that a similar tunneling procedure corresponding to atom-pair tunneling has already been observed for few-atoms confined  
in a driven double-well in \cite{Chen}. 
However, for the quenched system the tunneling takes place only within the SP mode, while 
the remaining tunneling modes are supressed, due to the quench,  
even for $\omega_{\rm{D}}\in\Delta \omega_{\rm{D}_1}$. To illustrate the effect of an interaction quench upon the driven lattice, on the tunneling dynamics, Figure 3(b) shows the probability $A_1(t)$ both 
for the unquenched and the quenched system for various driving frequencies. 
As it can be observed the effect of the quench depends on the driving frequency. Indeed, for $\omega_{\rm{D}}\leq \min(\Delta\omega_{\rm{D}_1})$ the quench decreases the frequency of the tunneling branch (see the 
red empty circles in Figure 3(a) which correspond to the interaction quenched fidelity spectrum) and leads to a significant enhancement of the amplitude of this tunneling branch (e.g. see the blue and 
black line in Figure 3(b)). The latter is a consequence of the fact that the 
interaction quench injects energy to the system. However, for $\omega_{\rm{D}}>\max(\Delta\omega_{\rm{D}_1})$ the tunneling branch is quite insensitive to the quench by means that 
both the frequency and the amplitude of the tunneling 
probability are slightly larger (see Figures 3(a) and (b)).  

To determine the frequencies of the local dipole mode in the outer wells we
calculate the spectrum $\Delta {\rho _L}(\omega ) =
\frac{1}{\pi }\int {dt} \Delta {\rho _L}(t){e^{i\omega t}}$.
The analysis of the corresponding breathing component will be performed in the next subsection, where we shall examine in more detail the effects 
of the quench dynamics. Figure 4(a) presents $\Delta\rho_{L}(\omega)$, where two 
emergent frequency branches (denoted as ($a_1$) and ($a_2$) in the spectrum) of the intra-well oscillations   
are visible. It is observed that for driving frequencies $\omega_{\rm{D}}\in [0,1.5]$ the intra-well dipole mode possesses  
two distinct frequencies which come into resonance in the region $\omega_{\rm{D}} \in[2,3]$ and then for $\omega_{\rm{D}}>3.0$ are again well separated.  
To gain insight into the impact of an interaction quench, performed on top of the driving, on the intra-well density oscillations, 
Figure 4(b) shows $\Delta \rho_L(t)$, at resonance ($\omega_{\rm{D}}=2.875$) for different quench amplitudes, 
namely at $\delta g=0.0, 1.0$ and $2.0$. As expected (resonance) $\Delta \rho_L(t)$ features a beating dynamics but with an increasingly decaying envelope with increasing  
quench amplitude, which is a direct effect of the interactions. A similar dephasing behavior holds for the other $\omega_{\rm{D}}$'s where $\Delta \rho_L(t)$ 
does not exhibit a beating pattern. Concerning the width of the resonant 
region different amplitudes of the interaction quench    
lead to a slight broadening of the resonant region. 
According to our calculations for the case with $\delta g=0$ the resonant frequency region 
corresponds to $\omega_{\rm{D}} \in [2,3]$, while for $\delta g=1.0$ and $\delta g=2.0$ the corresponding regions are $\omega_{\rm{D}} \in [1.8,3.2]$ and $\omega_{\rm{D}} \in [1.5,3.5]$ respectively.  
Summarizing, one can induce this resonant intra-well dynamics by
adjusting the driving frequency and by applying an interaction quench to increase the width of the resonance and manipulate the amplitude of the intra-well oscillations.

From another perspective the above-mentioned resonant behavior can be illustrated by employing the occupation of the 
zeroth band of the triple well during the evolution. 
The probability of finding all the four bosons within the zeroth band (employing the multiband 
expansion) reads   
\begin{equation}
|B_{\{N_i\};\{I_i\}}(t)|^2=\sum_{\{I_i\}} |\left\langle
N_{1}^{(I_1)},N_{2}^{(I_2)},N_{3}^{(I_3)}| \Psi(t) \right\rangle|^2, 
\end{equation}
where the summation is performed over the excitation  
indices with the imposed constraints 
$\sum_{i=1}^3 n_i^{(1)}=N$ and $\sum_{i=1}^3 \sum_{j=2}^3 n_i^{(j)}=0$ (see also Eq.(3)). 
Figure 4(c) shows the probability $|B_{\{N_i\};\{I_i\}}(t)|^2$ for all the bosons to
reside in the zeroth band for various driving frequencies
$\omega_{D}$ and a fixed amplitude $\delta=0.03$.   
At resonance a complete depopulation of the
zeroth band at some specific time intervals is observed. To be more precise, this 
probability exhibits a revival-like behavior on short time-scales, and decays as time evolves (see in particular the black dashed curve in Figure 4(c)).   
The local minima of $|B_{\{N_i\};\{I_i\}}(t)|^2$ are  
connected to the enhancement of the amplitude of the oscillations of the single-particle
density (see also Appendix B). On the other hand, for driving frequencies
away from $\Delta \omega_{\rm{D}_1}$ the respective probability for all the bosons to
occupy the zeroth band is rather large and is indeed 
dominant. However, significant contributions e.g. at $\omega_{\rm{D}}=5.25$ or $\omega_{\rm{D}}=9.25$ (see Figure 4(c)) from excited configurations cannot
be neglected, especially in the regions $\Delta\omega_{\rm{D}_1}$, $\Delta\omega_{\rm{D}_2}$ 
where the system departs from the initial state (see also Figure 1(a)) in a prominent way. 
Finally, in order to explore the impact of the interaction quench at resonance, Figure 4(d) shows $|B_{\{N_i\};\{I_i\}}(t)|^2$ for
different quench amplitudes at $\omega_{D}=2.75$. It is observed that 
for larger interaction quenches, this probability exhibits a more strongly decaying envelope which  
is a pure effect of the interactions. As it can be seen for increasing
quench amplitude the probability for the system to remain in the zeroth band, in the course of the dynamics, 
decays on increasingly shorter time scales and the system is dominated by different types of 
excitations, e.g. two, three or four particles distributed in the first and second excited bands, as expected intuitively.

\subsection{Case II: Periodically driven dynamics for different interaction quench amplitudes}

In the following, we shall examine the impact of the   
quench amplitude $\delta{g}$, focusing on two different driving frequency regions, i.e.   
for an almost adiabatic 
periodic driving and in the vicinity of the resonance (see also Figure 1(a)).   
To obtain an overview of the dynamical response, Figures 5(a) and (b) show the fidelity evolution with respect to $\delta{g}$, for 
fixed driving frequencies $\omega_{\rm{D}}=0.75$ and $\omega_{\rm{D}}=2.75$ respectively. As it is expected, for larger quench 
amplitudes the time-evolved final state deviates from the initial (ground) state in 
a prominent way. For instance, $\bar{F}_{\{\omega_{\rm{D}}=0.75,\delta g=0\}}=0.95$ and $\bar{F}_{\{\omega_{\rm{D}}=0.75,\delta g=4.0\}}=0.6$, while $\bar{F}_{\{\omega_{\rm{D}}=2.75,\delta g=0\}}=0.7$ and 
$\bar{F}_{\{\omega_{\rm{D}}=2.75,\delta g=4.0\}}=0.4$. 
Next, let us proceed with a more detailed analysis in order to probe the effect of an  
interaction quench on the inter-well tunneling dynamics and the intra-well excited modes. 

To examine the tunneling dynamics, Figure 6(a) presents the fidelity spectrum $F_{\{\delta g\}}(\omega)=\frac{1}{\pi} \int dt F_{\{\delta g\}}(t)$   
as a function of the quench amplitude. Three inter-well tunneling branches ($a_1'-a_3'$) can be identified. The lowest branch ($a_1'$),  
which dominates for strong quench amplitudes, refers to the energy difference $\Delta\epsilon$ within the energetically lowest-band  
states of the SP mode, e.g. from the initial state $\ket{1^{(0)},2^{(0)},1^{(0)}}$ to a final 
state $\ket{2^{(0)},1^{(0)},1^{(0)}}$ etc. The second branch ($a_2'$) corresponds to tunneling between the SP and DP   
modes, e.g. from $\ket{1^{(0)},2^{(0)},1^{(0)}}$ to   
$\ket{2^{(0)},2^{(0)},0^{(0)}}$ etc. The third branch ($a_3'$) refers to a tunneling 
process among the SP and T modes, e.g. from $\ket{1^{(0)},2^{(0)},1^{(0)}}$ to    
$\ket{3^{(0)},1^{(0)},0^{(0)}}$ etc. The remaining inter-well tunneling branches 
which correspond to transitions of energetically higher different modes are negligible in comparison to the aforementioned and therefore we can hardly identify them in Figure 6(a).  
To probe the effect of the driving frequency on the tunneling spectrum, Figure 6(b) shows $F_{\{\delta g\}}(\omega)$ at $\omega_{\rm{D}}=2.75$ (i.e. at resonance of 
an explicitly driven triple well) with varying quench amplitude.  
The three observed tunneling branches ($b_1'-b_3'$) refer to the same transitions, i.e. between the same number states as addressed above, but they are slightly shifted 
to higher frequencies as a consequence of the higher driving frequency. The remaining branches, e.g. $a_4'$ and $b_4'$, that are visible in the spectrum, which show  
more prominent deviations for the different driving frequencies correspond to other modes 
and inter-band transitions and will be explained below. 

To identify the frequencies of the local breathing mode we resort to the second moment 
$\sigma_i^2(\omega ) = \frac{1}{\pi}\int{dt\sigma_i^2(t){e^{i\omega t}}}$ for each well (see Sec. II.B and Eq.(5)). Focusing on the 
left well, which possesses a breathing component (see also Figure 2(b)) we calculate  
the frequency spectrum of $\sigma_L^2(\omega )$ which matches the branch $a_4'$ in the 
fidelity spectrum (see Figure 6(a)). Most importantly this frequency branch resonates with two distinct tunneling branches 
at different quench amplitudes, namely at $\delta g\approx1.0$ with the branch $a_3'$ (see the elipse in Figure 6(a)) and at $\delta g\approx2.8$ with the 
branch $a_2'$ (see the dashed elipse in Figure 6(a)) of the tunneling. 
Turning to the middle well, Figure 6(c)   
presents $\sigma_M^2(\omega )$, thus showing  
two main peaks ($a_1''$-$a_2''$) with respect to the quench amplitude.  
The lowest of these peaks refers to a tunneling mode (see also Figure 6(a)) being identified from the energy
difference within the energetically lowest states of the SP 
mode. The appearance of this peak in the spectrum is due to the fact
that the tunneling can induce a modulation on the width of the local
wavepacket. The second peak located at $\omega_{2} \approx 4.5$ refers to an inter-band process,  
i.e. to a transition from $\ket{1^{(0)},2^{(0)},1^{(0)}}$ to
$\ket{1^{(0)},1^{(0)}\otimes 1^{(2)},1^{(0)}}$.
Inspecting now more carefully the fidelity spectrum in Figure 6(a) we observe that the latter breathing frequency branch $a_2''$ (being denoted as $a_5'$ in Figure 6(a))  
comes into a resonance with the highest tunneling frequency branch ($a_3'$) at large quench 
amplitudes $\delta g \approx 5.2$. However, this tunneling branch is not visible in Figure 6(c) due to its small amplitude in comparison to the breathing ($a_2''$) branch.
To comment on the dependence of the breathing peak ($a_2''$) on the interaction quench we observe that it  
is more sensitive to $\delta g$ for $0.0<g_{f}<2.5$, otherwise it is approximately constant.
To probe the effect of the driving frequency on the breathing branch of the middle well, Figure 6(d) illustrates the 
spectrum of $\sigma_M^2(\omega)$ with respect to a varying $\delta g$ for $\omega_{\rm{D}}=2.75$. The respective breathing branches    
denoted as $b_1''$, $b_2''$ in the Figure, are slightly disturbed in comparison to the case with $\omega_{\rm{D}}=0.75$. Concerning the first one, we have already 
commented on its deviation in our discussion of Figures 6(a), (b). Focusing now on the highest frequency branch of the breathing a significant alteration is 
observed: for small quench amplitudes, $0.0<\delta g<0.8$, it possesses a single frequency, while for $\delta g>0.8$ the branch splits into two, with slightly different 
frequencies. The first is near to the corresponding frequency for $\omega_{\rm{D}}=0.75$ but slightly larger, while the second is larger than both. 
\begin{figure*}[ht]
        \centering
           \includegraphics[width=0.60\textwidth]{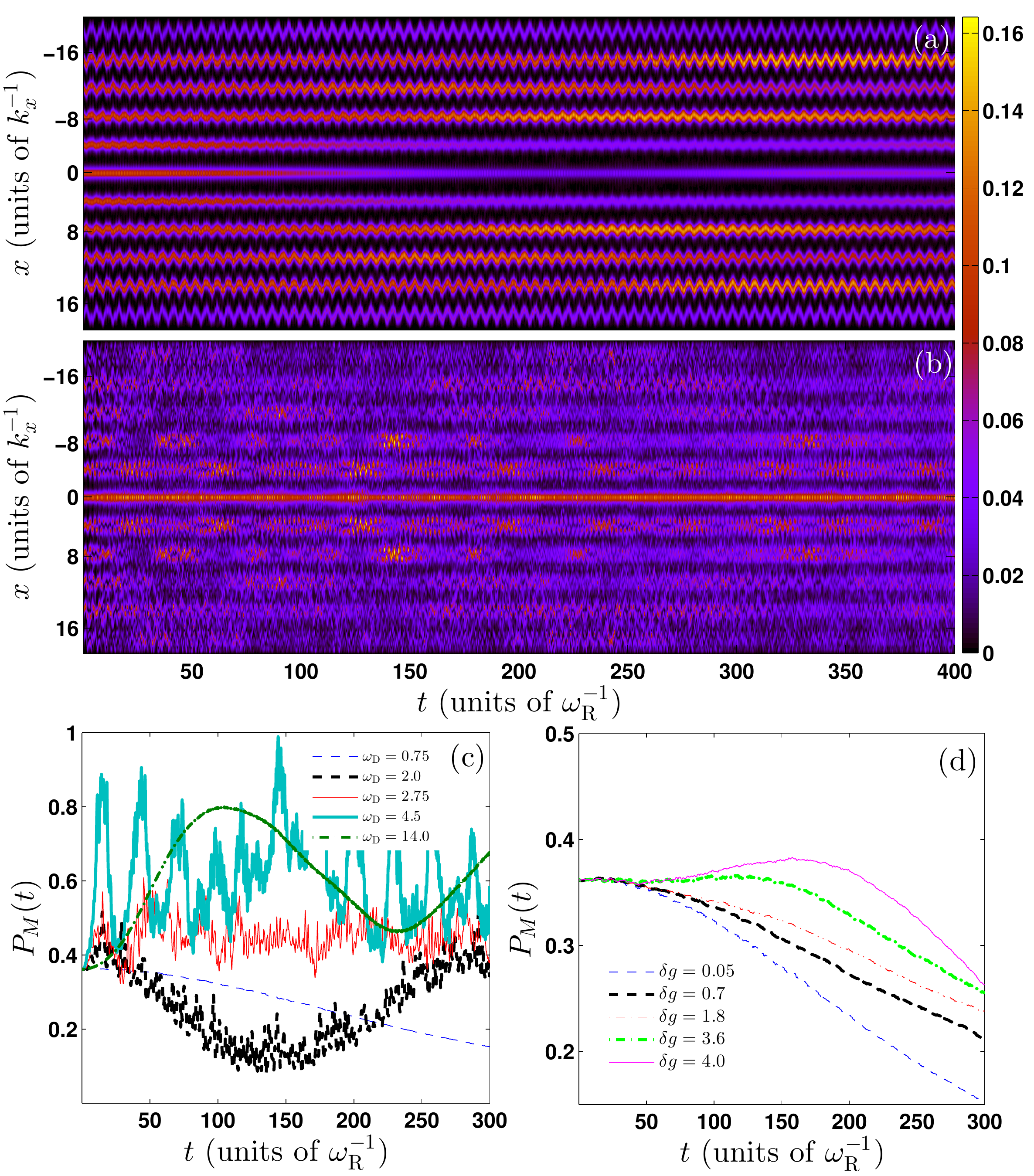}
                \caption{Time evolution of the one-body density $\rho_{1}(x,t)$ in a periodically driven eleven-well
                potential for different driving frequencies: (a) $\omega_{D}=1.25$ and (b) $\omega_{D}=2.875$. The
                driving amplitude is fixed to the value $\delta=0.03$, while the initial state corresponds to the ground state of five 
                weakly interacting bosons with $g=0.05$. (c) Probability of finding all the bosons in the central well ($P_M(t)$) during the evolution for different 
                driving frequencies $\omega_{\rm{D}}$ (see legend). (d) The same as (c) 
                but for $\omega_{\rm{D}}=0.75$ and different quench amplitudes $\delta g$ (see the legend).}
\end{figure*}
Finally, let us quantitatively examine the dipole component in the outer wells by employing 
the frequency spectrum $\Delta {\rho _L}(\omega ) = \frac{1}{\pi }\int {dt}
\Delta {\rho _L}(t){e^{i\omega t}}$ for various quench amplitudes. 
Figure 6(e) shows $\Delta\rho_L(\omega)$  
where we can identify three dominant peaks (denoted as $a_1'''-a_3'''$) which are 
located at ${\omega _1'''} \approx 1.2$, 
${\omega _3'''} \approx 2.5$, while $\omega_2'''$ is quench dependent. The steady frequency branches ($a_1'''$ and $a_3'''$)
correspond to the dipole mode and refer to inter-band transitions, e.g. from $\ket{1^{(0)},2^{(0)},1^{(0)}}$ to
$\ket{1^{(0)}\otimes 1^{(1)},1^{(0)},1^{(0)}}$ or to $\ket{1^{(0)}\otimes 1^{(2)},1^{(0)},1^{(0)}}$ respectively.   
On the other hand, the 
quench dependent frequency peak ($a_2'''$) is related to the third inter-well tunneling mode (being denoted in Figure 6(a) as $a_3'$). As shown in Figure 6(e) the latter  
branch $a_2'''$ experiences two resonances with each dipole branch at different quench amplitudes, namely at $\delta g \approx 0.7$ with the lowest frequency dipole branch ($a_1'''$) and 
at $\delta g \approx 3.0$ with the higher frequency dipole branch $a_3'''$. Moreover, by examining once more the fidelity spectrum (Figure 6(a)) more carefully, it is obsrved that   
the highest frequency dipole branch experiences a resonance with the second inter-well tunneling mode ($a_2'$) at $\delta g \approx5.0$.  
In order to conclude on the dependence of the dipole branches on the driving frequency we show in Figure 6(f) the $\Delta {\rho _L}(\omega)$ at $\omega_{\rm{D}}=2.75$.   
As shown the lower frequency dipole branch ($a_1'''$) is strongly dependent on the driving frequency (see branch $b_1'''$ in Figure 6(f)), while the 
higher frequency branch ($a_3'''$) is essentially unaffected. Most importantly, the aforementioned resonant 
behavior is still existing for $\omega_{\rm{D}}=2.75$ but in this case two more resonances appear in the spectrum (see Figure 6(b)) due to a shift of the lowest frequency dipole branch. These resonances are 
located at $\delta g\approx2.1$ and $\delta g\approx 4.0$ and refer to a coupling among the second ($b_2'$) and third ($b_3'$) tunneling branches with the lowest frequency dipole branch.  

In the next section, we proceed to the investigation of a system
with filling $\nu  < 1$ in order to generalize our findings. In
particular, by considering a setup with eleven wells and five particles  
we demonstrate that the above discussed resonant behavior for the intra-well dynamics induced by 
an explicitly driven potential is present also here. Subsequently, we explore the impact of an   
interaction quench.

\section{Quench dynamics in the driven lattice for filling factor $\nu<1$}

Here we shall concentrate on a larger lattice system characterized by a filling factor smaller 
than unity, namely we consider the case of five bosons trapped in an eleven-well potential. To understand and 
interpret the dynamics let us firstly briefly comment on the ground state properties of the system. An important 
property of the ground state is the spatial redistribution of the
atoms as the interparticle repulsion increases. The non-interacting ground state (g=0) is the product of the single-particle eigenstates spreading across 
the entire lattice, while the presence of the hard-wall boundaries render the neighborhood of the central well of the potential slightly more populated. Increasing the   
repulsion within the weak interaction regime the atoms are pushed to the outer sites which gain and lose population 
in the course of increasing $g$ \cite{Brouzos1}.

In the following, let us firstly focus on the driven bosonic dynamics induced, at $t=0$, by a vibrating eleven-well potential 
to the ground state of five repulsively interacting bosons with $g=0.05$. Figures 7(a), (b) demonstrate the response of the system 
on the one-body level for different driving frequencies $\omega_{\rm{D}}$, but the same driving amplitude $\delta=0.03$.   
The overall out-of-equilibrium behavior shows similar characteristics as
in the case of the triple well, i.e. the occurrence of out-of-phase dipole-like modes among the outer wells of the lattice, a local-breathing mode 
in the central well and an inter-well tunneling mode accompanying the dynamics. In addition, a transition from a non-resonant (Figure 7(a)) to a resonant 
intra-well dynamics (Figure 7(b)) by adjusting $\omega_{\rm{D}}$ is observed at the same frequency $\omega_{D}=2.875$ as in the 
triple well case. This resonant behavior is again manifested (Figure 7(b)) on the one-body density evolution as 
the formation of enhanced density oscillations at each site being further related to a gradual depopulation of the zeroth band during the evolution.
In terms of the significant contributing number states we can infer that out-of-resonance the dynamics can well be described 
by the set of lowest-band states (with a small contribution from the excited band states), while at resonance the inclusion 
of number states which obey the constraints $\sum_{i=1}^{11} n_i^{(1)}=N-1$, $n_i^{(3)}=0$ and $n_i^{(2)}=1$
for $k=1,...,11$ is necessary. Contributions from excited states to the second band, i.e.  
$\sum_{i=1}^{11} n_i^{(1)}=N-1$, $n_i^{(2)}=0$ and $n_i^{(3)}=1$ 
for $k=1,...,11$ also exist but they are negligible in comparison to the excitations to the first excited band.

Another important observation here is that by tuning the driving frequency $\omega_{\rm{D}}$ close to resonance 
the tunneling dynamics is modified. To explicate   
the latter, we employ as a measure of the inter-well tunneling the spatially integrated middle-well 
density $P_M(t)=\int_{-\pi/2}^{\pi/2} dx \rho_1(x,t)$, shown in Figure 7(c) for different 
driving frequencies, namely before, exactly at and after the resonance. Approaching $\omega_{\rm{D}}=2.875$ from below a diffusion to the outer wells is observed. 
At the region of $\omega_{\rm{D}}=2.875$ the tunneling dynamics is slowed down, i.e. the occupation of the middle well is fluctuating around a mean value. 
For $\omega_{\rm{D}}>2.875$ the tunneling process is modified and a tendency for the particles to concentrate in the central well is observed. 
Employing a corresponding number state analysis we can infer that for $\omega_{\rm{D}}>2.875$ states with higher occupancy in the central well gain prominence. The 
same behavior of the tunneling dynamics (before and after the resonance) is also observed in the triple well case. 
Furthermore, let us inspect the influence of an interaction quench on top of the driven lattice. As expected intuitively, by increasing 
the interaction quench the tunneling process decreases.  
Figure 7(d) shows $P_M(t)$ for different interaction quench 
amplitudes on top of the periodically driven lattice with $\omega_{\rm{D}}=0.75$ (i.e. away from the resonance). It is observed that $P_M(t)$ becomes steady for increasingly longer times     
as we increase $\delta g$, thus indicating a decrease of the corresponding inter-well tunneling dynamics.  
Finally, note that due to the low filling the admixing modes, induced after an interaction quench upon the periodically driven lattice, in the outer  
wells are hardly visible and therefore not shown here. 
\begin{figure}[ht]
        \centering
           \includegraphics[width=0.50\textwidth]{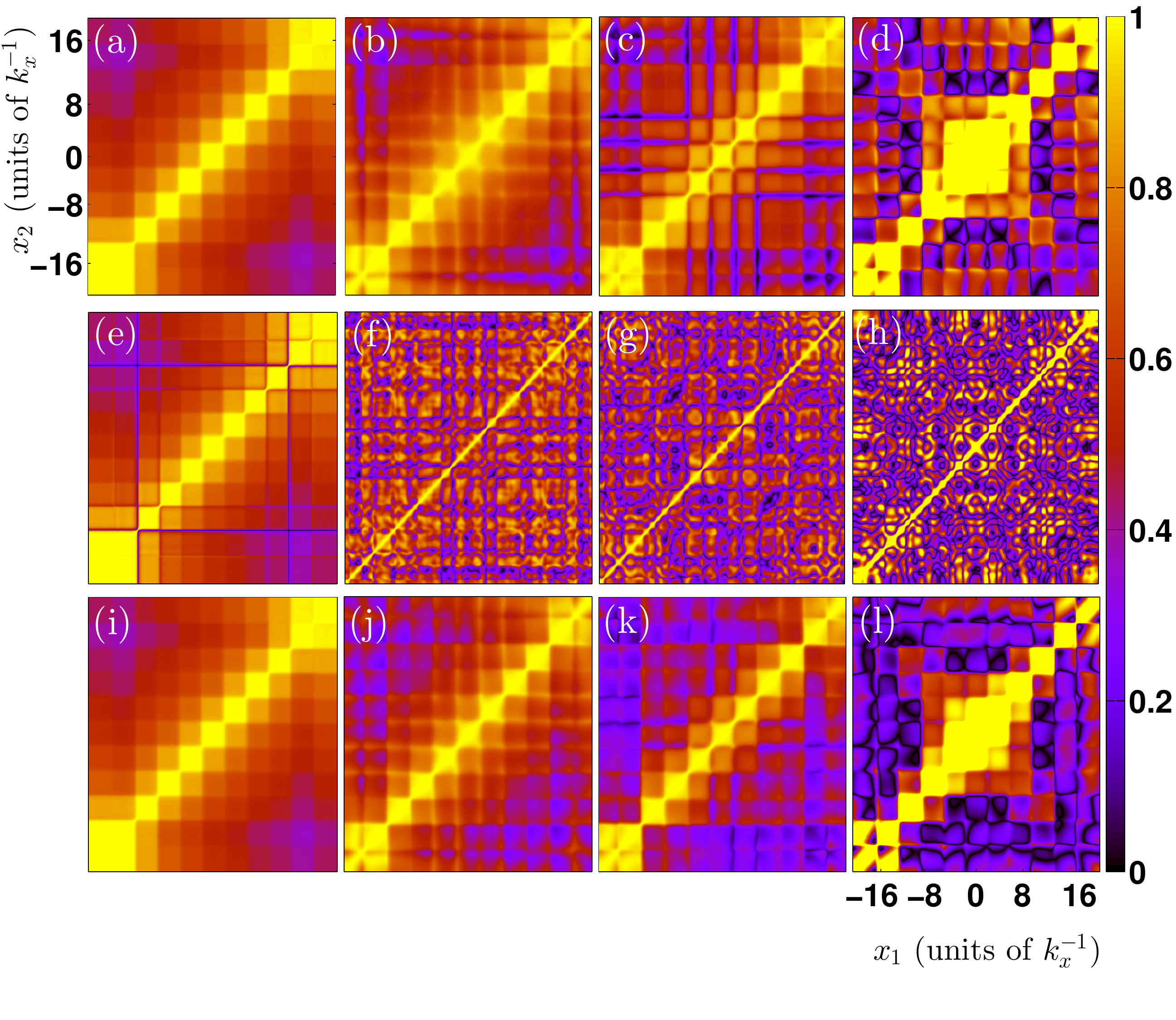}
                \caption{One body coherence function for different time instants ($t_1=1.0$, $t_2=56.0$, $t_3=123.0$ and $t_4=193.0$) during the evolution 
                caused by a periodically driven eleven-well potential (a,b,c,d)  
                with $\omega_{\rm{D}}=0.75$ and (e,f,g,h) $\omega_{\rm{D}}=3.0$. (i,j,k,l) shows the evolution of the one-body coherence in a periodically driven potential with 
                $\omega_{\rm{D}}=0.75$ and a simultaneous interaction quench with amplitude $\delta g=1.0$. The
                driving amplitude is fixed to the value $\delta=0.03$ and the initial state corresponds to the ground state of five 
                weakly interacting bosons with $g=0.05$.}
\end{figure}

Let us further investigate the signature of the resonant regions as well as the effect of the interaction 
quench on top of the periodically driven lattice by exploring the first order correlation function (see Eq.(8)), in 
coordinate space, which  
quantifies the degree of spatial coherence of the interacting system \cite{Naraschewski}. It is 
important to stress that, within the single-orbital Gross-Pitaevskii theory, the quantum wavepacket remains coherent 
at all times in contrast to a many-body calculation where it exhibits prominent time-varying 
structures which in turn indicate the rise of fragmentation on the system as the  
correlations between particles build up. From this point of view we expect a strong influence on  
the change of the spatial distribution of the atoms in the lattice either due to the 
resonant driving or as a consequence of the interaction quench.  
Foccussing on small driving frequencies ($\omega_{\rm{D}}=0.75$)
within the weakly interacting regime ($g=0.05$) we observe the spread of the coherence (Figures 8(a),(b),(c),(d)) through the lattice sites as time evolves.   
The diagonal elements are always perfectly coherent and their first neighboors remain close to unity throughout  
the time evolution. The off-diagonal elements are partially coherent, and are  
oscillating around the value 0.5, while for comparatively long evolution times 
a site selective off-diagonal long range order appears (see Figure 8(d)).     
Turning our attention to the resonant driving (see Figures 8(e),(f),(g),(h)) a different behavior throughout  
the time evolution is observed: On short time scales only the diagonal elements remain coherent and the off-diagonal is partially coherent. As time evolves 
a substantial loss of coherence even on the diagonal is observed,    
while the off-diagonal elements exhibit a much more prominent and complex structure. A direct comparison at equal times 
of the correlation function for non-resonant and resonant driving shows that   
resonant driving and loss of coherence go hand in hand with each other. 
On the other hand, by performing an interaction quench on top of the driving, 
the coherence (see Figures 8(i),(j),(k),(l)) is unity along the diagonal, while for sufficiently long evolution times 
it tends to vanish away from the diagonal.   
Finally, note that the off-diagonal contributions tend to fade out (but never vanish completely even for stronger quenches since the particles 
always remain delocalized) with increasing quench amplitude 
and a tendency for concentration close to
the diagonal is observed at equal times. This indicates that   
the strength of the interaction between the particles strongly affects the 
correlations; the stronger the inter-particle repulsion, the stronger the loss of coherence.
As a concluding remark we can infer that either the resonant driving or a 
quench on top of the driving entails an intensified loss of coherence.

\section{Conclusions and Outlook}

In the present work, the few-body correlated non-equilibrium quantum dynamics of an interaction quenched bosonic cloud   
in an external periodically driven finite-size optical lattice has been
investigated. The effect of an interaction quench on top of the driven  
lattice has been analyzed. We focus on large lattice depths and small driving amplitudes in
order to limit the degree of excitations that could lead to the creation 
of the cradle motion \cite{Mistakidis1} or even to
heating processes. Starting from the ground state of a weakly 
interacting small atomic ensemble, we examine in detail the
time evolution of the system in the periodically driven 
optical lattice by a simultaneous interaction quench.  

It has been shown that for the case of the periodically driven lattice  
one can induce out-of-phase local dipole modes in the outer wells, while a local breathing mode can be generated in the central 
well. This is in direct contrast with a shaken lattice, where only in-phase dipole modes are excited.   
A wide range of driving frequencies has been considered in order to unravel the range from adiabatic to high  
frequency driving. 
We observe that within the intermediate frequency regimes, being intractable by current analytical methods, the system can be driven to a 
far out-of-equilibrium state when compared to other driving frequency regions.
In particular, a resonance of the intra-well dynamics occurs with enhanced tunneling dynamics, thus opening energetically 
higher-lying inter-well tunneling channels.
A prominent signature of the resonant regions as well as the effect of the 
interaction is provided via the study of the time-dependence of the first order coherence, where   
intensified loss of coherence is observed. This loss of coherence constitutes 
an independent signature of the resonant regions, allowing to study it 
from another perspective and, potentially, to measure it in experiments.
Following an interaction quench on top of the periodically driven lattice for various driving frequencies, we can trigger more effectively the inter-well as well as 
the intra-well dynamics and steer the system towards strongly out-of-equilibrium regimes. Here, the tunneling as well as the local breathing mode 
in the middle-well are amplified, while in the outer wells the atomic cloud experiences an admixture of a dipole and a breathing component.
This admixture leads to simultaneous oscillations around the minimum of the well as well as a    
contraction and expansion in the course of the dynamics.  
Our analysis shows that one can use the interaction quench to manipulate the tunneling frequency 
rendering the single-particle tunneling dominant even at resonance. 
Concerning the on-site modes it is shown that an interaction quench can be used in order to manipulate their amplitude oscillations 
yielding also a strong influence on the excitation dynamics.

Subsequently, the dynamics of the periodically driven lattice (i.e. for a fixed driving frequency) as a function of the quench amplitude has been studied.  In particular, the tunneling contains three  
modes, the breathing possesses two frequency branches and the corresponding admixture three branches: one from the breathing component and 
two which refer to the dipole component. Furthermore, five resonances 
between the inter-well tunneling dynamics and the intra-well dynamics have been revealed.  
The inter-well tunneling experiences a resonance with the breathing component of the central well, 
two resonances with the breathing component of the outer wells and two resonances with the dipole component of the 
outer wells. These resonances can further be manipulated 
via the frequency of the periodic driving. As a result, the combination of different driving protocols 
can excite different inter- and intra-well modes as well as manifest various energetically higher components of a mode. Most importantly, 
the observed resonances between different inter- and intra-well modes demonstrate the richness of the system, while their 
dependence on various system parameters, e.g. the driving frequency shows the tunability of the system.  
The above-mentioned realization of multiple resonances constitutes arguably one of the central results of our investigation, which 
to the best of our knowledge has never been reported in such a setting.

Finally, let us comment on possible future extensions of the present work. Our analysis 
reveals that a combination of different driving protocols can induce admixtures 
of excited modes which in the present case corespond to admixtures of dipole-like and breathing-like modes.  
In this direction, it would be a natural next step to find the optimal pulse of the interaction quench protocol in order to induce 
a perfectly shaped squeezed 
state. Also the understanding and prediction of the long-time dynamics imposing the 
interaction quench on the driven lattice at different transient times is certainly of interest.

\appendix

\section{Harmonic oscillator: Admixtures of dipole-like and breathing-like modes}

In the present Appendix we shall briefly demonstrate the creation of admixtures of excitations consisting of a dipole and a breathing component in the dynamics of 
a bosonic ensemble confined in a one-dimensional harmonic oscillator. To begin with, let us firstly comment on the creation 
of each of the above excited modes separately. It is well known that a quench on the frequency of the 
harmonic oscillator or on the interatomic repulsive interaction induces a breathing mode oscillation of the atomic cloud. On the other hand, a sudden 
displacement or a periodic driving, e.g. shaking, of the harmonic oscillator can induce a dipole mode in the atomic 
cloud. However, a combination of the above techniques can induce in the dynamics \cite{Streltsova} more complicated modes and 
requires computational methods 
which can take into account higher orbitals, i.e. correlations. Here, we aim at illuninating this scenario by examining the evolution of an atomic 
cloud consisting of six bosons initially ($t<0$) prepared in the ground state of an harmonic oscillator potential. Subsequently,  
($t>0$) the cloud is subjected to a periodic driving and a simultaneous quench on the interatomic repulsive interaction. 
Thus, the Hamiltonian that governs the dynamics reads
\begin{equation}
\label{eq:2}H = \sum\limits_{i = 1}^N \bigg({ \frac{{{p_i ^2}}}{{2M}}}+ {V_{D}}({x_i};t)\bigg) +
g_{f}\sum\limits_{i < j} {\delta({x_i} - {x_j})},
\end{equation}
where the periodic driving of the harmonic oscillator is modelled via the time-dependent 
potential $V_{D}(x;t)= \frac{\omega^2}{2}(x-A\sin(\omega_{D}t))^2$  
and $\delta{g}=g_{f}-g_{in}$ denotes the quench amplitude.
\begin{figure}[ht]
        \centering
           \includegraphics[width=0.45\textwidth]{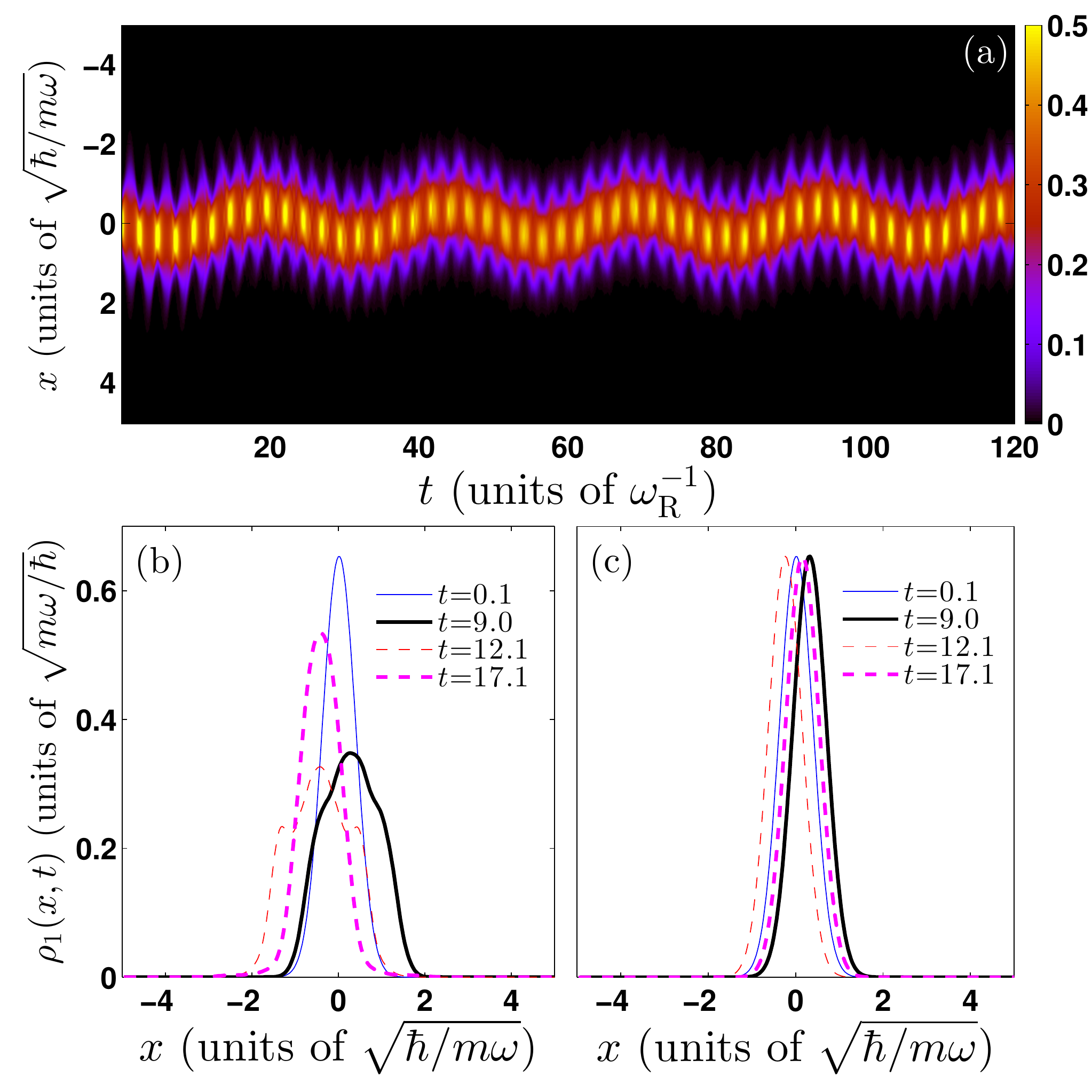}
                \caption{(a) Time evolution of the one-body density $\rho_{1}(x,t)$ caused by a periodically driven harmonic trap 
                with $\omega_{D}=0.25$ and a simultaneous interaction quench with amplitude $\delta g=1.6$. The
                driving amplitude is fixed to the value $A=0.6$, while the initial state corresponds to the ground state of six 
                weakly interacting bosons with $g=0.05$. We also illustrate the one-body density profiles at certain time instants (see legend) during the 
                evolution of the periodically driven oscillator with (b) $\delta g=1.6$ and (c) $\delta g=0$. }
\end{figure}
Figure 9(a) illustrates the dynamics of the atomic cloud on the single-particle level by employing the one-body 
density. It is observed that 
the cloud not only oscillates inside the external trap but also changes its shape during the 
oscillation. This is a clear signature that 
the induced mode is different from a pure dipole mode or a pure breathing mode but it is an admixture of the above mentioned excitations. 
To indicate explicitly this fact we illustrate in Figure 9(b) the profiles of the one-body density at certain time-instants   
during the evolution. The cloud compresses and decompresses (caused by 
the interaction quench) during its oscillation (caused by the driven oscillator) inside the external harmonic trap. On the contrary, a cloud 
which is only subjected 
to the above external driving (see Figure 9(c)) performs the well-known dipole oscillation and the wavepacket exhibits oscillations with 
constant width and amplitude.

\section{Remarks on the resonant intra-well dynamics of the driven lattice}

In the present Appendix we shall briefly comment on the characteristics of the resonant dynamics of the driven lattice 
from a one-body perspective. Indeed, Figure 10(a) 
presents $\rho_1(x,t)$ at $\omega_{\rm{D}}=2.75$. The overall dynamics exhibits enhanced density 
modulations being manifest as internal fast oscillations 
and large amplitude oscillations in each well of period $\sim 20$. The inter-well tunneling is also enhanced in comparison to small 
$\omega_{\rm{D}}$'s (see Figure 2(a)).
A similar intrawell resonant behavior has been observed in \cite{Mistakidis2}, where enhanced and in-phase 
oscillating dipoles have been revealed. On the contrary, here, we observe enhanced and out-of-phase oscillating dipole modes as well as an amplified 
breathing mode in the center. Thus, exploiting the presently used driving scheme we have the possibility to open an additional 
energetic channel.   
To quantify that the driven lattice induced dynamical features are independent of the interaction strength $g$ or the particle number $N$ 
we calculate the deviation of the local density oscillation from its mean value, i.e. $\Lambda=\int_0^Tdt|\Delta{\rho_{\alpha}(t)-\overline{\Delta\rho_{\alpha}}}|/T$, 
where $\overline{\Delta\rho_{\alpha}}=\int_0^Tdt\Delta\rho_{\alpha}(t)/T$ denotes 
the mean oscillation amplitude over the considered propagation time $T$ and $\Delta\rho_{\alpha}(t)$ refers to the intra-well wavepacket asymmetry.   
Figure 10(b) shows the mean amplitude of the intra-well oscillation for the left well as a function of the 
driving frequency $\omega_{D}$ for different interaction 
strengths $g$ but the same particle number. The mean 
amplitude with a varying $\omega_{D}$ increases up to $\omega_{D}=2.875$ where it exhibits a 
peak (position of the resonance) and then decreases 
again exhibiting several smaller peaks 
at frequencies where the system is driven far from equilibrium (see also Figure 1(a)).  
Comparing the dynamics for different interactions it is observed 
that the ensemble exhibits the same overall behavior 
but the mean oscillation amplitude is slightly larger (for higher interactions) especially in the region of the 
central peak. This is a direct interaction effect, since the system possesses more 
energy. On the other hand, in order to investigate 
whether the above results are independent of the particle number the same quantity ($\Lambda$) is shown in Figure 10(c) for 
varying particle number, namely $N=4,8$. The mean amplitude presents the 
same overall behavior with respect to 
the driving frequency $\omega_{D}$ but it is also slightly larger for increasing particle number with a maximal deviation of the order of 
$30\%$. 
\begin{figure}[ht]
        \centering
           \includegraphics[width=0.40\textwidth]{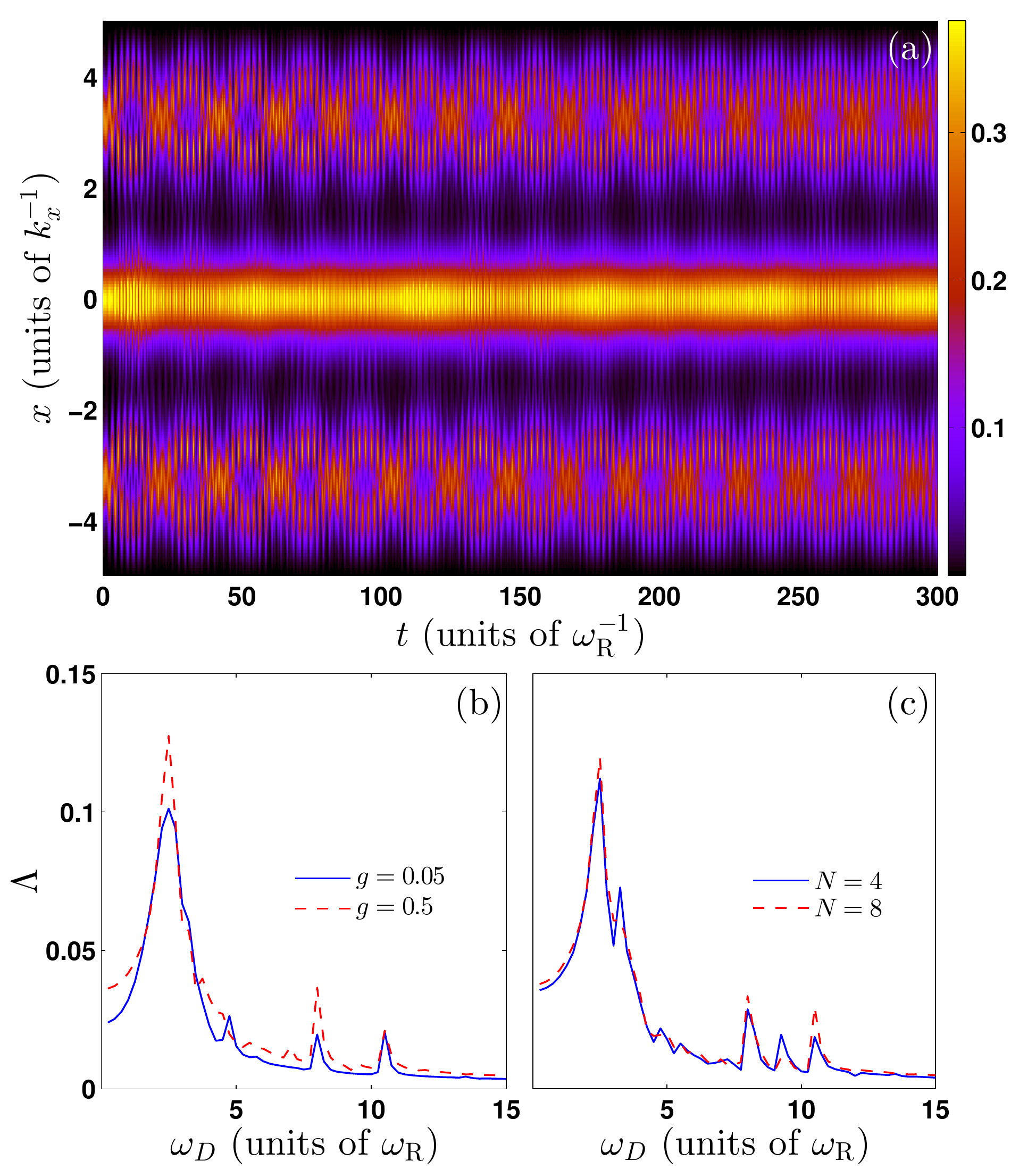}
                \caption{(a) Time evolution of the one-body density $\rho_{1}(x,t)$ in a triple well 
                for $\omega_{D}=2.75$. The
                driving amplitude is fixed to the value $\delta=0.03$, while the initial state corresponds to the ground state of four
                weakly interacting bosons with $g=0.05$. (b) Mean oscillation amplitude $\Lambda$ of the left well for $N=4$ bosons 
                as a function of the driving 
                frequency $\omega_{\rm{D}}$ for different interparticle repulsion (see legend). (c) The same as (b) but  
                for fixed interaction $g=0.2$ and different particle number (see legend).}
\end{figure}

\begin{figure}[ht]
        \centering
           \includegraphics[width=0.50\textwidth]{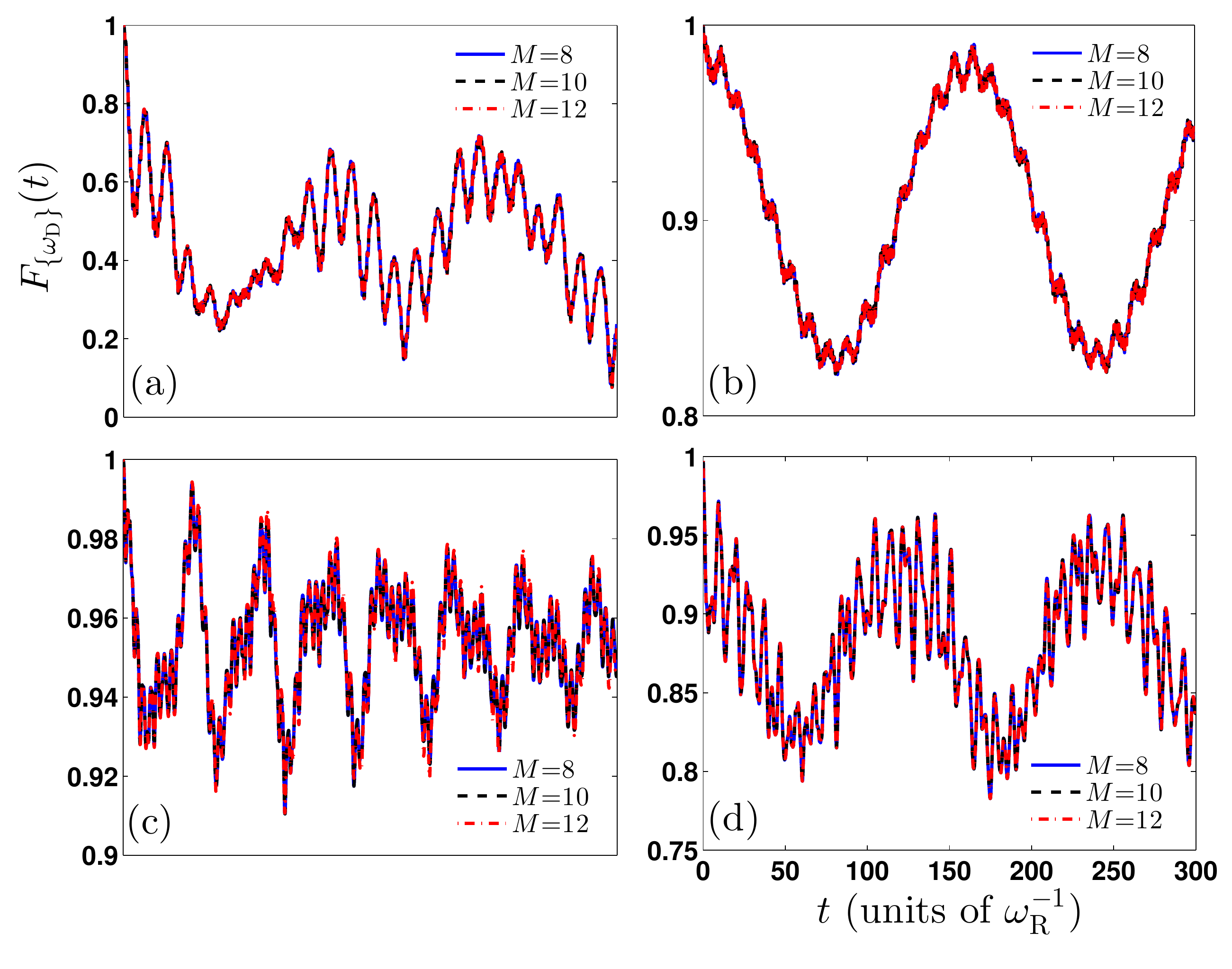}
                \caption{Fidelity evolution $F_{\omega_{\rm{D}}}(t)$ of a periodically driven triple well with (a) $\omega_{\rm{D}}=2.5$ and 
                (b) $\omega_{\rm{D}}=7.5$ with an increasing number of SPFs (see legend). (c), (d) $F_{\omega_{\rm{D}}}(t)$ for various SPFs (see legend) with a simultaneous interaction 
                quench of amplitude (c) $\delta g=0.5$ and (d) $\delta g=2.0$ on top of the periodically driven triple well with $\omega_{\rm{D}}=0.75$. }
\end{figure}

\begin{figure}[ht]
        \centering
           \includegraphics[width=0.50\textwidth]{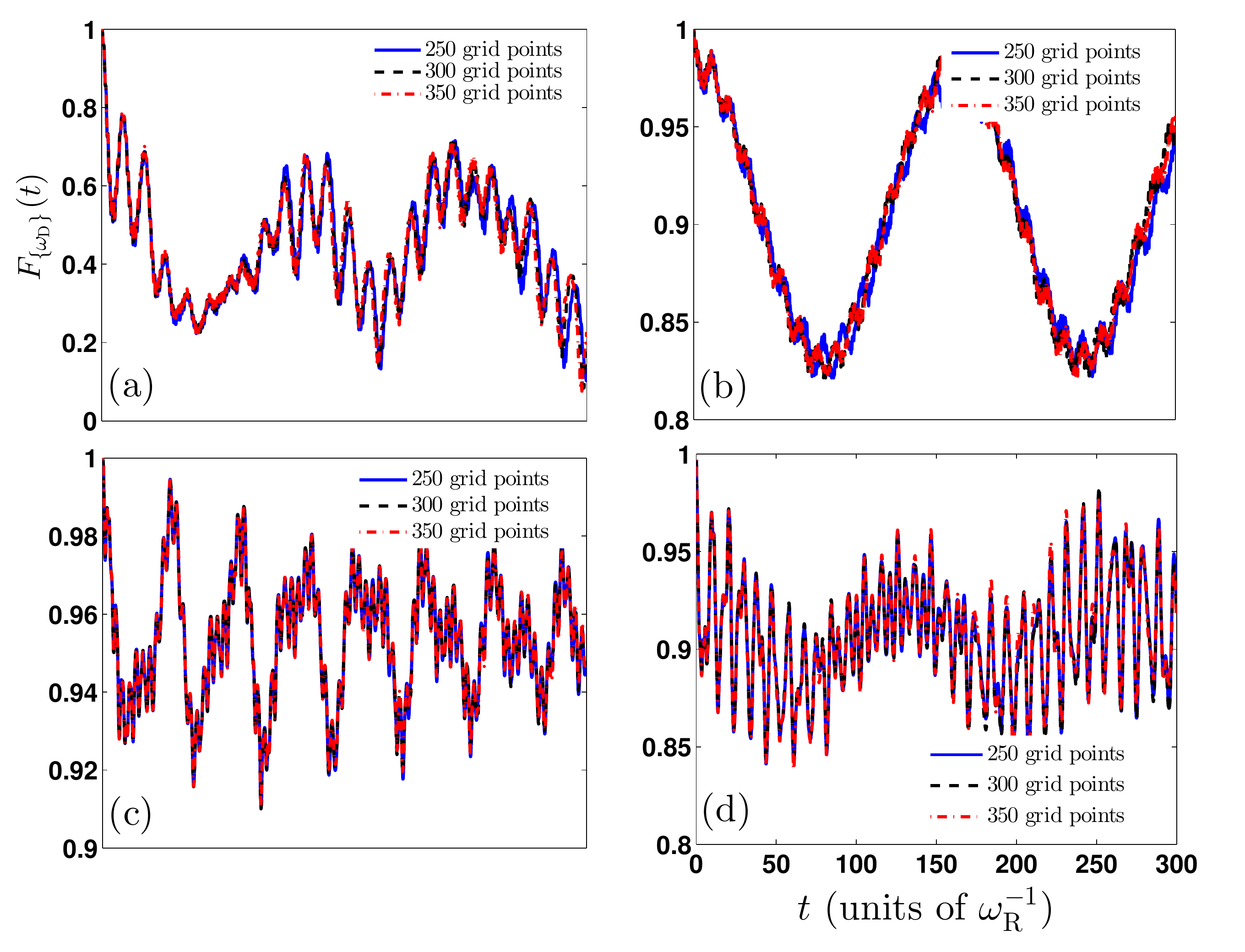}
                \caption{Fidelity evolution $F_{\omega_{\rm{D}}}(t)$ of a periodically driven triple well with (a) $\omega_{\rm{D}}=2.5$ and 
                (b) $\omega_{\rm{D}}=7.5$ with an increasing number of grid sizes (see legend). (c), (d) $F_{\omega_{\rm{D}}}(t)$ for various grid sizes (see legend) with a simultaneous interaction 
                quench of amplitude (c) $\delta g=0.5$ and (d) $\delta g=2.0$ on top of the periodically driven triple well with $\omega_{\rm{D}}=0.75$.}
\end{figure}

\section{The Computational Approach: MCTDHB}

To solve the many-body Schr\"{o}dinger equation $\left( {i\hbar {\partial _t} - H} \right)\Psi (x,t) = 0$ 
of the interacting bosons as an initial value problem $\ket{{\Psi (0)}} = \left| {{\Psi _0}}
\right\rangle$, we employ the Multi-Configuration
Time-Dependent Hartree method for Bosons (MCTDHB) \cite{Alon,Alon1,Streltsov}. 
The latter constitutes an efficient and accurate method both for the 
stationary properties and the non-equilibrium quantum
dynamics of systems consisting of a single bosonic species  
and has already been applied for a wide set of 
problems, see e.g. \cite{Streltsov,Streltsov1,Alon2,Alon3}.  
The wavefunction is represented by a set of
variationally optimized time-dependent orbitals which implies
an optimal truncation of the Hilbert space by employing a
time-dependent moving basis where the system can be
instantaneously optimally represented by time-dependent permanents.
Thus, the many-body wavefunction which is expanded in terms of
the bosonic number states $\left| {{n_1},{n_2},...,{n_M};t}
\right\rangle$, based on time-dependent single-particle
functions (SPFs) $\left| \phi_{i}(t) \right\rangle$, $i=1,2,...,M$,
reads
\begin{equation}
\label{eq:10}\left| {\Psi (t)} \right\rangle  = \sum\limits_{\vec n
} {{C_{\vec n }}(t)\left| {{n_1},{n_2},...,{n_M};t} \right\rangle }.
\end{equation}
Here $M$ is the number of SPFs and the summation $\vec n$ is over
all the possible combinations $n_{i}$ such that the total number
of bosons $N$ is conserved. Note that in the limit in which $M$ approaches the number
of grid points the above expansion is equivalent to a full configuration interaction
approach. However, in the case of $M=1$ the many-body wavefunction is given by a single 
permanent $\ket{n_{1}=N;t}$ and the method reduces to the time-dependent Gross Pitevskii equation. 
To determine the time-dependent wave function $\left|
\Psi(t) \right\rangle$ we need the equations of motion for the
coefficients ${{C_{\vec n }}(t)}$ and of the SPFs $\left| \phi_{i}(t) \right\rangle$.
Following e.g. the Dirac-Frenkel
\cite{Frenkel,Dirac} variational principle, i.e. ${\bra{\delta
\Psi}}{i{\partial _t} - \hat{ H}\ket{\Psi }}=0$, we end up with the
well-known MCTDHB equations of motion
\cite{Alon,Streltsov,Alon1} consisting of a set of $M$
non-linear integrodifferential equations of motion for the orbitals
which are coupled to the $\frac{(N+M-1)!}{N!(M-1)!}$ linear
equations of motion for the coefficients. Finally, let us remark that in terms of our implementation we 
use an extended version of MCTDHB being referred to in the literature as the Multi-Layer Multi-Configuration 
Time-Dependent Hartree method for Bosons (ML-MCTDHB) \cite{Cao,Kronke}. This 
package is particularly suitable for treating systems consisting of different bosonic species, while for the case 
of a single species it reduces to MCTDHB. 

For our numerical implementation a discrete variable representation (DVR) for the
SPFs and a sine-DVR, which intrinsically introduces hard-wall
boundaries at both edges of the potential, has been employed. The
preparation of the initial state has been performed by using the so-called
relaxation method in terms of which one obtains the lowest
eigenstates of the corresponding $m$-well setup. The key idea is to
propagate some trial wave function ${\Psi ^{(0)}}(x)$ by the
non-unitary operator ${e^{ - H\tau }}$. This is equivalent to an imaginary time
propagation and for $\tau  \to \infty $, the propagation converges to
the ground state, as all other contributions (i.e., $e^{-E_n\tau}$) are 
exponentially suppressed. In turn, we periodically drive the optical lattice and perform 
a quench on the strength of the interparticle repulsion and study the
evolution of $\Psi ({x_1},{x_2},..,{x_N};t)$ in the $m$-well
potential within MCTDHB.

Within our simulations the following overlap criteria are fullfilled
$\left| \langle \Psi | \Psi \rangle - 1 \right| < 10^{-9}$ and 
$\left| \langle \varphi_i | \varphi_j \rangle - \delta_{ij} \right| < 10^{-10}$ for the total 
wavefunction and the SPFs respectively.
Furthermore, to ensure the convergence of our simulations we have used up to 12(11) optimized single
particle functions for the triple-(eleven) well, thereby observing a systematic convergence of our
results for sufficiently large spatial grids. In particular, we have used 350 spatial grid points in the case of a  
triple-well and 800 spatial grid points for the eleven-well potential.  
In the following, let us briefly demonstrate the convergence procedure concerning our simulations either with an increasing number of    
SPFs (and fixed number of 350 grid points) or for a varying number of grid points and a fixed number 
of SPFs, $M=12$. Figure 11 shows the fidelity evolution for different numbers of SPFs, namely $M=8,10,12$, for the driven triple well at driving 
frequencies $\omega_{\rm{D}}=2.5$, $\omega_{\rm{D}}=7.5$ (see Figures 11(a), (b) respectively) and $F_{\omega_{\rm{D}}}(t)$ by  
employing simultaneous interaction quenches with amplitudes $\delta g=0.5$, $\delta g=2.0$ on top of the driving, $\omega_{\rm{D}}=0.75$, (see Figures 11(c), (d) respectively). A systematic convergence 
of the fidelity evolution (for $M>8$) is observed for increasing number of SPFs. 
For instance, the maximum deviation (at $\omega_{\rm{D}}=2.5$) observed in the fidelity evolution (see Figure 11(a)) calculated using 8 and 12 SPFs respectively is of the 
order of $0.3\%$ at large evolution times ($t>200$).
Furthermore, in order to show the convergence with an increasing number of  
grid points Figure 12 presents the fidelity evolution of the driven triple well at $\omega_{\rm{D}}=2.5$, $\omega_{\rm{D}}=7.5$ (see Figures 12(a), (b) respectively)  
and by performing interaction quenches with $\delta g=0.5$, $\delta g=2.0$ on top of the driven triple well, $\omega_{\rm{D}}=0.75$, (see Figures 12(c), (d) respectively). Again, we 
observe convergence for an increasing number of grid points (especially for grid sizes that contain larger than 300 spatial grid points). 
For instance, the maximum deviation (at $\omega_{\rm{D}}=2.5$) observed in the fidelity evolution (see Figure 12(a)) calculated using 300 and 350 grid points respectively (and 12 SPFs) is of the 
order of $0.1\%$ at large evolution times ($t>250$). The same analysis has also been performed for the 
eleven well case (omitted here for brevity) showing the same behavior.  
Another criterion that confirms the achieved convergence is the population of the 
lowest occupied natural orbital kept in each case below $0.1\%$.

\section*{Acknowledgments}
The authors gratefully acknowledge funding by the Deutsche Forschungsgemeinschaft (DFG) in the framework of the
SFB 925 ''Light induced dynamics and control of correlated quantum
systems''. The authors thank C.V. Morfonios for fruitful discussions.

{}

\end{document}